\documentclass{aa_2024}  
\bibpunct{(}{)}{;}{a}{}{,}

\usepackage{graphicx}
\usepackage{float}
\usepackage{stfloats}
\usepackage{caption}
\usepackage{subcaption}
\usepackage{mathrsfs}
\usepackage{comment}
\usepackage{txfonts}
\usepackage{hyperref}
\hypersetup{colorlinks = true, linkcolor=red, citecolor=blue, urlcolor=blue}
\usepackage{natbib}
\usepackage{siunitx}
\usepackage {indentfirst}
\usepackage{longtable}
\usepackage{booktabs}
\usepackage{caption}
\begin{document} 
   \def\ergs{erg\,s$^{-1}$}
   \def\msunyr{M$_{\odot}$\,yr$^{-1}$}
   \def\kms{\mbox{km~s$^{-1}$}}
   \def\mgiidoub{\ion{Mg}{ii}\,$\lambda\lambda$\,$2795,\,2802$}

     \title{Revisiting atmospheric Roche lobe overflow in symbiotic binaries}
   \authorrunning{T. Liu et al.}

   \author{Tan Liu\inst{1,2,3},
           Natalia Ivanova\inst{4},
           Philipp Podsiadlowski\inst{5,6},
           Joanna Miko\l{}ajewska\inst{7},
           Zheng-Wei Liu\inst{1,2},
           Xuefei Chen\inst{1,2,7},
           Xiangcun Meng\inst{1,2},
           Zhanwen Han\inst{1,2,7}
          }

   \institute{Yunnan Observatories, Chinese Academy of Sciences (CAS), Kunming 650216, P.~R.~China\\
              \email{zwliu@ynao.ac.cn}
        \and
            International Centre of Supernovae (ICESUN), Yunnan Key Laboratory of Supernova Research, Kunming 650216, P.~R.~China
        \and 
            University of Chinese Academy of Sciences, Beijing 100049, P.~R.~China
        \and
            Department of Physics, University of Alberta, Edmonton, T6G 2E7, Alberta, Canada
        \and
            London Institute of Stellar Astrophysics, Vauxhall, London, United Kingdom
        \and
            University of Oxford, St Edmund Hall, Oxford OX1 4AR, United Kingdom
        \and
            Nicolaus Copernicus Astronomical Center, Polish Academy of Sciences, Bartycka 18, 00–716 Warsaw, Poland
        \and
            Key Laboratory for the Structure and Evolution of Celestial Objects, CAS, Kunming 650216, P.~R.~China
             }

   \date{Received xxx xx, 202x; accepted xxx xx, 202x}

  \abstract
  {Classical binary evolution models predict dynamically unstable mass transfer in symbiotic stars with high mass ratios ($q_\mathrm{crit} > 1.6$), leading to a common envelope (CE). However, a large fraction of observed S-type symbiotic systems maintain long-lived interaction, suggesting that an additional stabilizing mechanism may be at work.}
  {We investigate whether atmospheric Roche lobe overflow (RLOF) can prolong the mass transfer phase in symbiotic binaries and thereby reconcile theoretical predictions with observed system numbers.}
  {We implement the Rapid Unified Mass Transfer (RUMT) framework within \texttt{MESA} and compute a grid of white dwarf - giant binaries across a wide parameter space in donor mass, mass ratio, and orbital period. We then compare the resulting lifetimes and evolutionary tracks with a sample of well constrained Galactic S-type symbiotic systems.}
  {For convective giant donors, our models recover dynamically stable mass transfer up to $q\simeq1.5$, consistent with previous studies, but atmospheric overflow strongly extends the symbiotic phase. RGB and early-AGB systems with $q \lesssim 1.5$ can remain interacting for up to $10^6$ yr at $\dot M \gtrsim 10^{-9}\,M_\odot\,\mathrm{yr}^{-1}$, two to three orders of magnitude longer than the commonly assumed in population studies $\sim10^3$ yr pre–CE lifetimes. In these models the orbit shrinks mildly or re-expands after mass-ratio is less than 1.0. Systems with higher mass ratios still evolve toward a CE, but even for $q\simeq2-4$ the symbiotic phase lasts $10^4-10^5$ yr, one to two orders of magnitude longer than usually assumed. The synthetic distribution in the $P$--$q$ plane and individual evolutionary tracks are consistent with observed S-type symbiotic binaries, including recurrent novae.}
  {The RUMT framework, which incorporates atmospheric RLOF, provides an explanation for the long-term stability of many symbiotic binaries and may account for their high observed occurrence rate.}
  

   \maketitle 
   \nolinenumbers

\section{Introduction}\label{sec:introduction}
\noindent More than 50\% of stars are in binary systems 
\citep{sana2012binary,2022A&A...667A..44G}. Symbiotic stars (SySts) are interacting binaries in which a white dwarf (WD) or rarely also a neutron star (NS) accretes matter from an evolved red giant either on the red giant branch (RGB) or asymptotic giant branch (AGB). They are characterized by high accretion rates, $\dot{M} \gtrsim 10^{-9}\,M_{\odot}\,\mathrm{yr}^{-1}$, necessary to detect the WD aside an evolved red giant donor \citep[e.g.,][]{1986syst.book.....K}.
Like many other interacting binaries SySts are usually defined by spectroscopic characteristics \citep{1984PASA....5..369A,1986syst.book.....K,2000A&AS..146..407B}, i.e. (1) a red continuum with absorption features of a late-type red giant; (2) strong emission lines of H I and He I and either additional lines with an ionizational potential of at least 30 eV (e.g. He II, [OIII], [Ne V],
and [Fe VII])  or an A- or F-type continuum with additional absorption lines from HI and He I
and singly ionized metals (the latter corresponds to a SySt in outburst); (3) the presence of  the Raman scattered OVI emission features, even if there is no absorption features of the cool star.

\par Based on their near infrared colours, SySts are mainly classified into two classes: S-type (Stellar - as they show the presence of cool stellar photospheres) and D-type (Dusty - as they indicate the presence of warm dust shells) \citep{webster1975symbiotic}. S-type SySts have 
donors that are typically normal giants, with orbital periods ranging from approximately 200 to 1000 days, accounting for about $80\%$ of all SySts \citep{2003ASPC..303....9M}. In contrast, D-type SySts are characterised by donors that are Mira variables surrounded by a dust envelope. They typically have much longer orbital periods, typically above 40 years. In D-type systems, large binary separations and long orbital periods are required to accommodate the extended envelope and 
dust shell of the evolved giant.

\par Figure~1 of \cite{2012BaltA..21....5M} illustrates the distribution of orbital periods derived from photometric and spectroscopic survey databases  \citep[e.g.,][]{brandi2005spectroscopic,brandi2009spectroscopic,2013AcA....63..405G,2015AJ....150...48F,2017AJ....153...35F}. It shows that the observed orbital periods $P$ of S-type WD SySts peak at 
approximately 600 days, and fewer than 30\% of systems have $P$ greater than 1000 days. 

\par This observational distribution stands in clear conflict with the predictions of existing population-synthesis models (PSM) of WD SySts, which have consistently failed to reproduce the observed orbital period distribution \citep{yungelson1995model,2006MNRAS.372.1389L,2012MNRAS.424.2265L}. 
In particular, the theoretical results of \cite{2006MNRAS.372.1389L} predict that the orbital periods of S-type SySts should peak near 1500 days, with only about 20\% of systems having $P < 1000$ days. 
These models also produce mass ratios (defined as the ratio of the donor mass $M_{\rm d}$ to the WD accretor mass $M_{\rm a}$) systematically lower than the observed values (see their Figure~10). 
Resolving this discrepancy requires a more advanced and physically consistent treatment of mass transfer than what has been adopted in previous PSM simulations.

\par A key question for understanding the population of SySts, in particular their number in the Galaxy, is the lifetime of the symbiotic phase, which depends on the nature of the mass transfer phase. This can involve stable mass transfer through Roche-lobe overflow (RLOF), traditional wind mass transfer \citep{whitelock1987symbiotic,paczynski1971evolutionary,starrfield1987review,prvsa2005computational}, or wind Roche-lobe overflow \citep{podsiadlowski2007origin}. The lifetime is then mainly determined by the nuclear evolution of the giant ($\sim 10^6$--$10^7$\,yr). Many S-type SySts contain red giants that are close to filling their Roche-lobes as indicated by their ellipsoidal light curve variations \citep[e.g.,][see also Table \ref{tab:observation}]{2003ASPC..303....9M,2012BaltA..21....5M,2007BaltA..16...49R,2013AcA....63..405G}. This variability is caused by tidal distortion. When the donor radius $R_{\rm d}$ approaches the Roche lobe radius $R_{\rm RL}$, tidal forces stretch the stellar atmosphere. This leads to variations in the visible surface area, causing the observed magnitude to vary periodically. Such variability indicates that RLOF is likely to play an important role in the mass transfer process of SySts.

\par However, the mass ratio $q$  for the observed S-type SySts typically ranges from 2 to 4 (see Table \ref{tab:observation}). According to binary evolution theory, systems with a mass ratio above 
the critical value $q_\mathrm{\rm crit}$  are expected to undergo dynamically unstable mass transfer, leading to a common-envelope (CE) phase \citep{ivanova2013common}. 
For donors similar to those observed in SySts -- that is, giants with deep convective envelopes and masses $\lesssim 3,M_\odot$ -- the critical mass ratio for the onset of dynamical instability is $q_{\rm crit} \approx 1.6$ if mass transfer is assumed to be fully conservative \citep{2015MNRAS.449.4415P}.
Therefore we may be seeing SySts in the phase just before the beginning of the CE phase. 

\par The pre-CE phase is generally believed to be a short-lived phase, possibly as short as $10^3$\,yr \citep{podsiadlowski2007origin,2010AIPC.1314...59C}. The number of observable SySts is then given by the rate of forming close WD+RGB binaries, $\nu_{\rm WD+RGB}$, times the time $r_{\rm symb}$ during which they appear as SySts. In their detailed binary population synthesis study, \cite{han1995formation} obtained a MW $\nu_{\rm WD+RGB}$ of $0.02-0.10$\,$\rm yr^{-1}$ (see their Table 4). Given by that not all formed WD-RGB binaries must pass through a symbiotic-like observable phase, less than a hundred SySts could be observed in the MW.

According to the most complete New Online Database of Symbiotic Variables\footnote{\url{https://sirrah.troja.mff.cuni.cz/~merc2025A&A...695A..61M/nodsv/}}  \citep{2019RNAAS...3...28M}, there are 284 confirmed Galactic SySts of which $\sim$210 ($\sim$75\%) are S-type SySts with WD accretors, i.e. those that are the targets of our present study. Based on position and velocity data of these known SySts \cite{Laversveiler_2025} estimated a minimum number of $800-4100$ for the total Galactic population of SySts, and this is the only so far published empirical limit. \cite{Laversveiler_2025} do not distinguish between S- and D-type nor between WD and NS accretors so their limit should be reduced to $\sim 600-3000$ of S-type SySts with WD accretors.  All of these estimates are strongly affected by selection effects, and hence are quite uncertain, but the potential discrepancy between theoretically expected numbers (if mass transfer is on the verge of becoming unstable) and observed numbers highlights the necessity to investigate the nature of the mass transfer process in SySts more closely.

\par To better reconcile theoretical predictions with observations, several refinements to the mass transfer prescriptions in SySt models have been introduced to reduce the predicted instability and extend the duration of the mass transfer phase. Standard models of mass transfer assume that the process begins when the donor star’s radius equals its Roche lobe radius, effectively treating the stellar surface as a sharp boundary. The volume-equivalent radius of Roche lobe is given by the approximation in \citet{1983ApJ...268..368E}:

\begin{equation}
R_{\mathrm{RL}} = 
\frac{0.49\, q^{2/3}}
     {0.6\, q^{2/3} + \ln\!\left(1+q^{1/3}\right)} \, a ,
\label{eq:rochelobe}
\end{equation}

\noindent where $R_{\rm RL}$ is the Roche lobe radius  and $a$ is the binary separation. In this formalism, mass transfer begins only when 
\mbox{$R_{\mathrm{d}} > R_{\mathrm{RL}}$}.

However, RGB donors possess extended atmospheres, allowing mass transfer to begin even before the photosphere reaches the Roche lobe. This phenomenon is known as {\it atmospheric Roche lobe overflow}  (atmospheric overflow). A theoretical framework for this process was developed by \citet{1988A&A...202...93R}, who modeled both the stellar atmosphere and the atmosphere-driven stream as isothermal. In this model, an external energy source is required to accelerate the flow to sonic velocity while maintaining a constant temperature from the donor’s photosphere to the $L_{\rm 1}$ point.

\begin{equation}
- \dot{M}_{\mathrm{d}} = 
\frac{1}{\sqrt{e}} \,
\rho_{\mathrm{ph}} \, v_{\mathrm{s}} \, Q \,
\exp\!\left( -\frac{R_{\mathrm{RL}} - R_{\mathrm{d}}}{H_{\mathrm{P}}} \right),
\label{eq:mdot}
\end{equation}

\noindent Here,  $R_{\rm d}$ is the  photospheric radius of the donor, $Q$ is the effective cross-sectional area of the flow in $L_{\rm 1}$ plane, $v_s$ is the isothermal sound speed, $\rho_{\mathrm{ph}}$ is the photospheric density, and $H_\mathrm{P}$ is the pressure scale height of the atmosphere. In addition to assuming an isothermal stream, several other simplifications are made. These relate to the density and entropy profiles of the stream as it passes through the $L_{1}$ point, as well as the shape of the Roche lobe equipotential near $L_{1}$. The approximation breaks down once the distance from $L_{1}$ along the $L_{1}$ plane becomes significant, i.e. for substantial overflow where the stream can no longer be treated as optically thin. A complementary optically thick prescription is provided by the formalism of \citet{1990A&A...236..385K} (hereafter KR), which assumes an adiabatic stream and employs a second-order Taylor expansion of the Roche potential near $L_1$. Another issue arises from the inconsistent treatment of atmospheric overflow and Roche lobe overflow as simultaneous processes. The two approaches are not mutually consistent when applied together.

To overcome these limitations, \citet{2024ApJ...971...64I} introduced a new method -- Rapid Unified Mass Transfer (RUMT) -- which unifies the treatment of atmospheric RLOF and ${L_1}$ stream outflow into a continuous mass transfer model. This approach incorporates three-dimensional (3D) binary properties derived from simulations \citep[e.g.,][]{2023ApJ...952..126P} and interpolates them into one-dimensional (1D) stellar evolution models, such as those implemented in \texttt{MESA} \citep{2011ApJS..192....3P, 2013ApJS..208....4P}. The resulting mass transfer rates differ significantly from those predicted by the analytic formalisms of \citet{1988A&A...202...93R} and \citet{1990A&A...236..385K}, with atmospheric mass loss rates up to an order of magnitude higher or two orders of magnitude lower depending on donor properties. 

    \citet{2010AIPC.1314...59C} were the first to investigate atmospheric overflow in SySts using the formalism developed by \citet{1988A&A...202...93R}, focusing on a limited set of binary configurations involving a $1.5M_\odot$ donor and a $0.75M_\odot$ companion across several orbital periods. Their results indicated that atmospheric overflow can significantly extend the mass transfer phase compared to models where mass transfer begins only after the donor fills its Roche lobe. 
    
    In this study, we extend this work to a large grid of models,  adopting the RUMT method  \citep{2024ApJ...971...64I}, implemented within \texttt{MESA}, to model  the mass transfer process in symbiotic binaries. Section~\ref{sec:methods} outlines our numerical methods for binary evolution. Section~\ref{sec:results} presents the main results and their implications. Section~\ref{sec:discussion} provides a detailed discussion, and our conclusions are summarized in Section~\ref{sec:conclusions}.

\section{Method and input parameters}
\label{sec:methods}
\subsection{Detailed binary evolution}
\noindent We used the \texttt{MESA} stellar evolution code (version 23.05.1) to model semi-detached WD+RG binaries \citep{2011ApJS..192....3P,2013ApJS..208....4P,2015ApJS..220...15P,2018ApJS..234...34P,2019ApJS..243...10P,2023ApJS..265...15J}. The binary evolution was computed with the \texttt{binary} module in \texttt{MESA}, which updates the orbital parameters together with mass and angular momentum transfer. Only the red-giant donor was evolved as a full stellar model. The white dwarf was modeled as a point-mass accretor: its gravity influenced the binary potential in which the donor evolved, and accreted material was added directly to the WD mass. The RUMT method was implemented through customized routines.

\par For the atmospheric boundary conditions in our simulations, we use the Eddington grey atmosphere as implemented in {\tt MESA}. In this approach, the surface boundary condition for the stellar model is obtained using the standard $T(\tau)$ relation, with the Rosseland mean opacity evaluated locally at each optical depth and the gravitational acceleration assumed constant throughout the atmosphere.

\par Our RUMT framework then uses this MESA-provided surface boundary condition as the starting point for reconstructing the detailed atmospheric structure above the stellar surface in order to determine the mass loss rate. In this reconstruction, unlike in the original MESA atmosphere, the gravitational acceleration is allowed to vary during the integration. Since most of the optical depth is accumulated in layers close to the stellar surface, we expect the use of varying rather than constant gravity to have only a limited effect on the surface boundary condition itself. By contrast, the mass loss rate is expected to be more sensitive to the structure of the outer, more extended atmospheric layers, where the effect of varying gravity becomes more important.\footnote{A more self-consistent treatment would be to determine the surface boundary condition using the modified gravity, rather than relying on the built-in {\tt MESA} atmospheric routine. Unfortunately, this would require a numerically intensive iterative procedure involving full atmospheric integrations at each step, as well as substantial modifications to MESA.}

\par The calculations were performed with the standard SDK associated with \texttt{MESA} version 23.05.1. The detailed choices for the equation of state, opacity tables, nuclear reaction network, mesh controls, and other numerical settings are specified in the inlists provided with the supplementary material. We did not impose a fixed number of mesh zones; instead, \texttt{MESA} adjusted the mesh automatically, with the number of zones varying substantially between models, from only about a hundred near the beginning of the evolution to several thousand in some low-mass donor models. These inlists, together with the customized routines, will be made publicly available through the MESA Zenodo community upon publication.

This study models RLOF in semi-detached WD+RG systems using the RUMT method, which accounts for both stream and atmospheric outflows in the total mass transfer rate:

\begin{equation}
\dot{M}_{\mathrm{tot}} =
\int_{0}^{r_{\mathrm{ph}}}
\rho_{\mathrm{L1}}(r_{l}) \,
c_{\mathrm{s,L1}}(r_{l}) \,
\mathcal{E}(r_{l}) \, dr_{l}
+ \int_{r_{\mathrm{ph}}}^{r_{\mathrm{out}}}
\rho_{\mathrm{at}}(r_{l}) \,
c_{\mathrm{s,at}}(r_{l}) \,
\mathcal{E}(r_{l}) \, dr_{l} \, .
\end{equation}

In the above, the integrations are carried out across the $L_1$ plane, where each $r_l$ corresponds to the semi-major axis of an ellipse formed by the intersection of the $L_1$ plane with a given three-dimensional equipotential surface. 
The lower limit of the first integral, $r_l = 0$, corresponds to the $L_1$ point. The upper limit, $r_{\mathrm{ph}}$, marks the location where the photosphere intersects the $L_1$ plane. The second integral extends from $r_{\mathrm{ph}}$ to $r_{\mathrm{out}}$, which defines the outer boundary of the atmosphere as imposed by the numerical setup.
$\rho_\mathrm{L1}(r_l)$ is the density of the donor’s gas as it reaches the $L_1$ plane, and $\rho_{\rm at}(r_l)$ is the density of the atmospheric gas as it reaches the $L_1$ plane beyond the photosphere.

The quantities $c_{\mathrm{s,L1}}(r_l)$ and $c_{\mathrm{s,at}}(r_l)$ are the local sound speeds of the gas at the $L_1$ plane, for the donor’s matter below the photosphere and in the atmosphere, respectively (this uses the conventional nozzle approximation, where the speed with which gas passes through the nozzle is equal to its local sonic velocity). The term $\mathcal{E}(r_l)$ denotes the circumference of the ellipse corresponding to the given equipotential circuit in the $L_1$ plane.   The 3D structure of the equipotentials and associated quantities is described in  tables published by \cite{2023ApJ...952..126P}. Properties of the atmosphere, such as density and temperature, are obtained using an Eddington grey atmosphere model with opacities that vary with local density and temperature. Gravity throughout the atmosphere is defined by the binary equipotential rather than assuming a spherically symmetric stellar potential.

In our calculations, we neglect mass loss due to the donor’s stellar wind. We adopt fully conservative mass transfer, meaning that all mass lost through Roche lobe overflow is accreted by the companion. This assumption is adopted because conservative mass transfer typically causes stronger orbital shrinkage and therefore produces shorter-lived symbiotic phases; it thus serves as a strict lower limit on the lifetime of SySts. 
Orbital evolution accounts only for the angular momentum loss associated with systemic mass loss, that is, material escaping from the binary system. Stellar winds are not activated in our models, so no additional mass or angular momentum loss through winds is included. The effects of gravitational radiation, magnetic braking, and tidal spin-orbit coupling are also neglected to isolate the role of mass transfer.
The simulation is stopped when the donor overflows the $L_2$ equipotential. 

\subsection{Grid of binary star parameters}\label{sec:binary grid parameter}

\noindent We constructed a grid of binary models to systematically investigate how different binary parameters affect the mass transfer process in SySts. An overview of the adopted parameter space is given in Table~\ref{table1}. The adopted parameter space follows the distribution estimated by \cite{2012BaltA..21....5M}. After removing systems in which the initial white dwarf mass exceeds the Chandrasekhar mass limit, the grid comprises 418 binary models. See Section~\ref{sec:timescale of the SySts} for additional details. The initial main sequence donor models were computed with a metallicity of $Z = 0.02$. The convective treatment is detailed in the inlists. For example, we used a mixing-length parameter of 2.0, with the \texttt{MESA} overshooting parameters \texttt{overshoot\_f = 0.15} and \texttt{overshoot\_f0 = 0.02}.

\renewcommand{\arraystretch}{1.5}
\begin{table}[htbp]
  \centering
  \caption{Table of parameter space for the simulation of SySt models. The initial donor mass $M_\mathrm{d,i}$ ranges from 1 $M_{\odot}$ to 3 $M_{\odot}$ in steps of 0.5 $M_{\odot}$, the initial mass ratio $q_\mathrm{i}$ ($M_\mathrm{d}/M_\mathrm{a}$) ranges from 1 to 5 in steps of 0.5, and the initial orbital period $P_\mathrm{i}$ spans from 100 days to 1100 days in steps of 100 days. }
  \label{table1}
  \begin{tabular}{cccccc}
    \hline
    \hline 
    $M_{\mathrm{d,i}}$ & step & $q_\mathrm{i}$ & step & $P_{\mathrm{i}}$ & step \\
    ($M_{\odot}$) & ($M_{\odot}$) &  &  & (days) & (days) \\
    \hline
    1.0 -- 3.0 & 0.5 & 1.0 -- 5.0 & 0.5 & 100 -- 1100 & 100 \\
    \hline
  \end{tabular}
\end{table}

\section{Results}
\label{sec:results}

\subsection{Different types of mass transfer}
\noindent Since emission lines must be visible against the strong continuum of the giant, we classify a system as a genuine symbiotic binary system only when the mass transfer rate satisfies $\dot{M} \geq 10^{-9}\,M_{\odot},\text{yr}^{-1}$. At lower rates, the accretor would be too faint to produce detectable luminosity and line emission \citep{1986syst.book.....K}. This threshold therefore sets the effective mass transfer timescale of observable SySts.

\par
\begin{figure}[!htbp]
\centering
\includegraphics[scale=0.35]{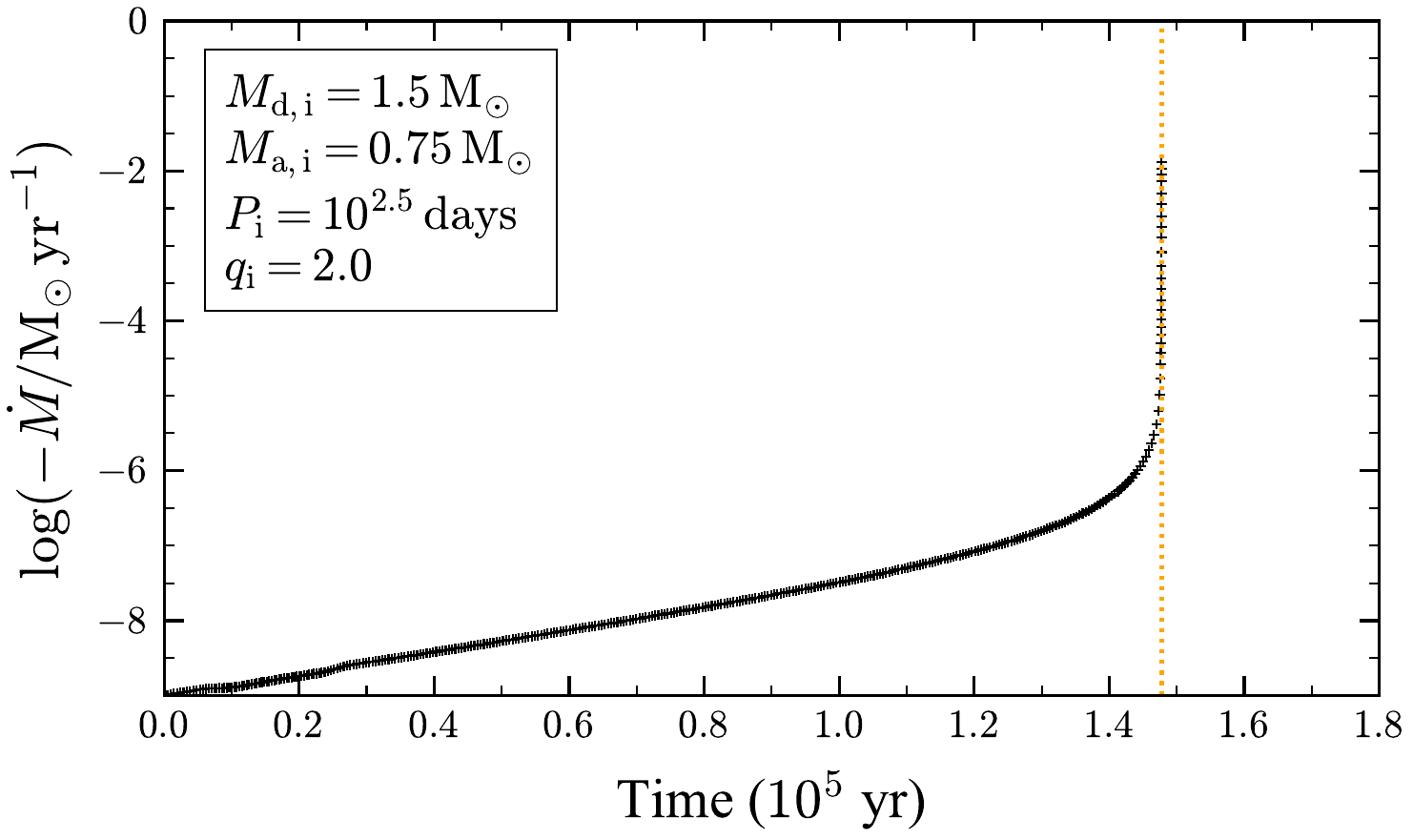}
\caption{An example of a model undergoing rapidly increasing (near-dynamical) mass transfer is shown. The initial donor mass is $1.5\,M_{\odot}$, the white dwarf accretor mass is $0.75\,M_{\odot}$, the initial mass ratio is $q_\mathrm{i}=2.0$, and the initial orbital period is $10^{2.5}$ days. At the onset of the SySt phase, defined as the moment when the mass transfer rate first exceeds $10^{-9}\,M_\odot\,\mathrm{yr}^{-1}$, the Roche-lobe filling factor ($R_\mathrm{d}/R_\mathrm{L}$) is 0.918. In this system, the mass transfer rate increases steeply and approaches the dynamical regime, eventually reaching the point where the donor overfills the $L_2$ equipotential. At that stage, the calculation is stopped, as a common-envelope phase is expected to follow but is not explicitly modeled. The duration over which the mass transfer rate exceeds $10^{-9}\,M_\odot\,\mathrm{yr}^{-1}$ is approximately $1.5\times10^5$ years. The orange dotted line marks the moment when the donor first exceeds the $L_2$ equipotential radius.}
\label{1}
\end{figure}

 In our calculations, we found that there are different types of mass transfer processes. Figure~\ref{1} shows an example of a model undergoing an unstable mass transfer process (exponentially increasing rate approaching the dynamical regime). Initially the binary system has a $1.5\,M_{\odot}$ giant and a $0.75\,M_{\odot}$ WD accretor. The initial orbital period is $10^{2.5}$ days. The origin of the x-axis corresponds to the time when the mass transfer rate exceeds $10^{-9}\,M_{\odot}\,\text{yr}^{-1}$. We found that at the onset on the SySts phase, the Roche lobe filling factor, defined as the ratio of the donor radius $R_\mathrm{d}$ to its Roche-lobe radius $R_\mathrm{L}$, is 0.918. For this system, the mass transfer rate rapidly increases and eventually enters the dynamical mass transfer phase, which may lead to a common envelope. The timescale during which the mass transfer rate exceeds $10^{-9}\,M_{\odot}\,\text{yr}^{-1}$ is approximately $1.5 \times 10^5$ yr.

\par
\begin{figure}[!htbp]
\centering
\includegraphics[scale=0.35]{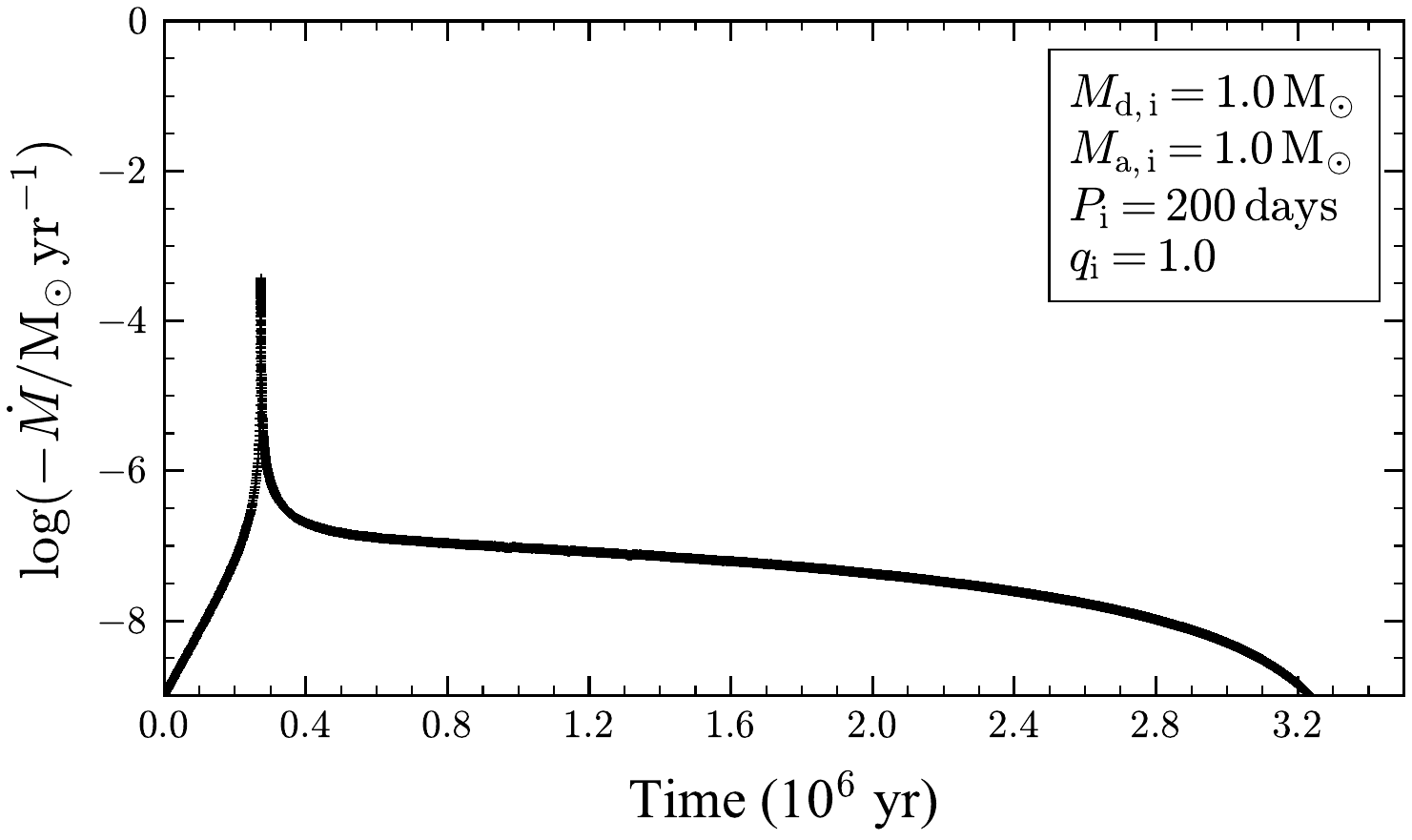}
\caption{An example of a model that experiences high but non-dynamical mass transfer is shown. The initial donor mass is $1.0\,M_{\odot}$, the white dwarf accretor mass is $1.0\,M_{\odot}$, the initial mass ratio is $q_\mathrm{i}=1.0$, and the initial orbital period is $200$ days. The mass transfer rate first rises to $\sim10^{-3}\,M_{\odot}\,\mathrm{yr}^{-1}$, but the system does not overflow the $L_2$ equipotential and no common envelope develops. Instead, the mass transfer rate subsequently declines and the system settles into a long-lasting, stable mass transfer phase. This phase lasts for approximately $3.2\times10^6$ years.
}
\label{2}
\end{figure}

 We also find models in which the mass transfer rate becomes very high but does not lead to unstable or dynamical mass transfer (Figure~\ref{2}). In the example shown, the initial donor mass $M_\mathrm{d,i}$ is 1.0 $M_{\odot}$, the initial WD accretor mass $M_\mathrm{a,i}$ is 1.0 $M_{\odot}$, the initial mass ratio $q_\mathrm{i}$ is 1.0 and the initial orbital period $P_\mathrm{i}$ is $200$ days. The mass transfer rate initially rises to approximately $10^{-3}\,M_{\odot}\,\mathrm{yr}^{-1}$ 
, but the system does not overfill the $L_2$ equipotential. Instead, the mass transfer rate subsequently decreases and the system settles into a stable mass transfer phase. The duration of this phase is approximately $3.2\times10^6$ years.
Similar evolutionary behaviour is found in other systems in the grid and will be discussed further in Section \ref{sec:timescale of the SySts}. The evolution of the mass transfer rate for all models is provided in the Appendix (see Figure~\ref{evolutionary_sequence}.).

\subsection{Timescale of SySts}\label{sec:timescale of the SySts}

\begin{figure*}[htbp]
    \centering
    \begin{minipage}{0.45\linewidth}
        \centering
        \includegraphics[width=0.9\linewidth]{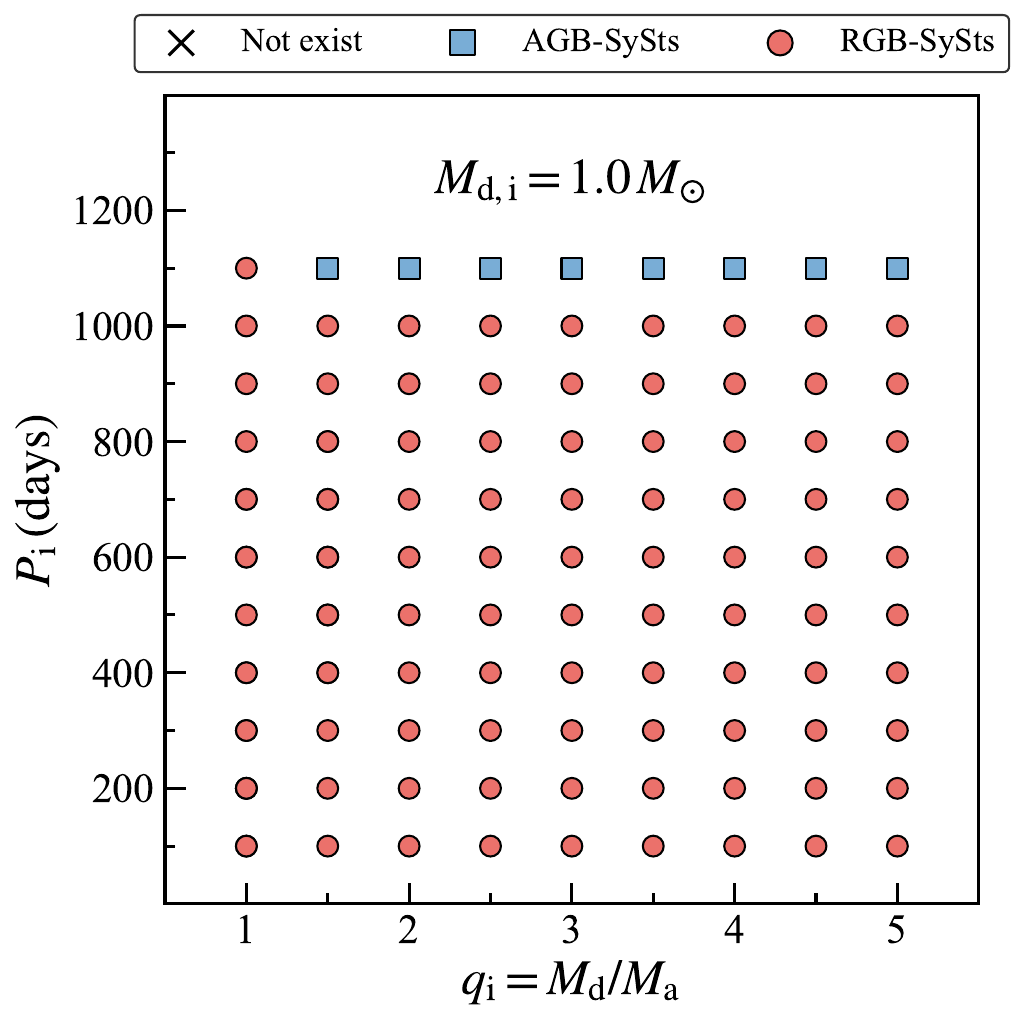}
        \label{fig:mass_1.0} 
    \end{minipage}
    \begin{minipage}{0.45\linewidth}
        \centering
        \includegraphics[width=0.9\linewidth]{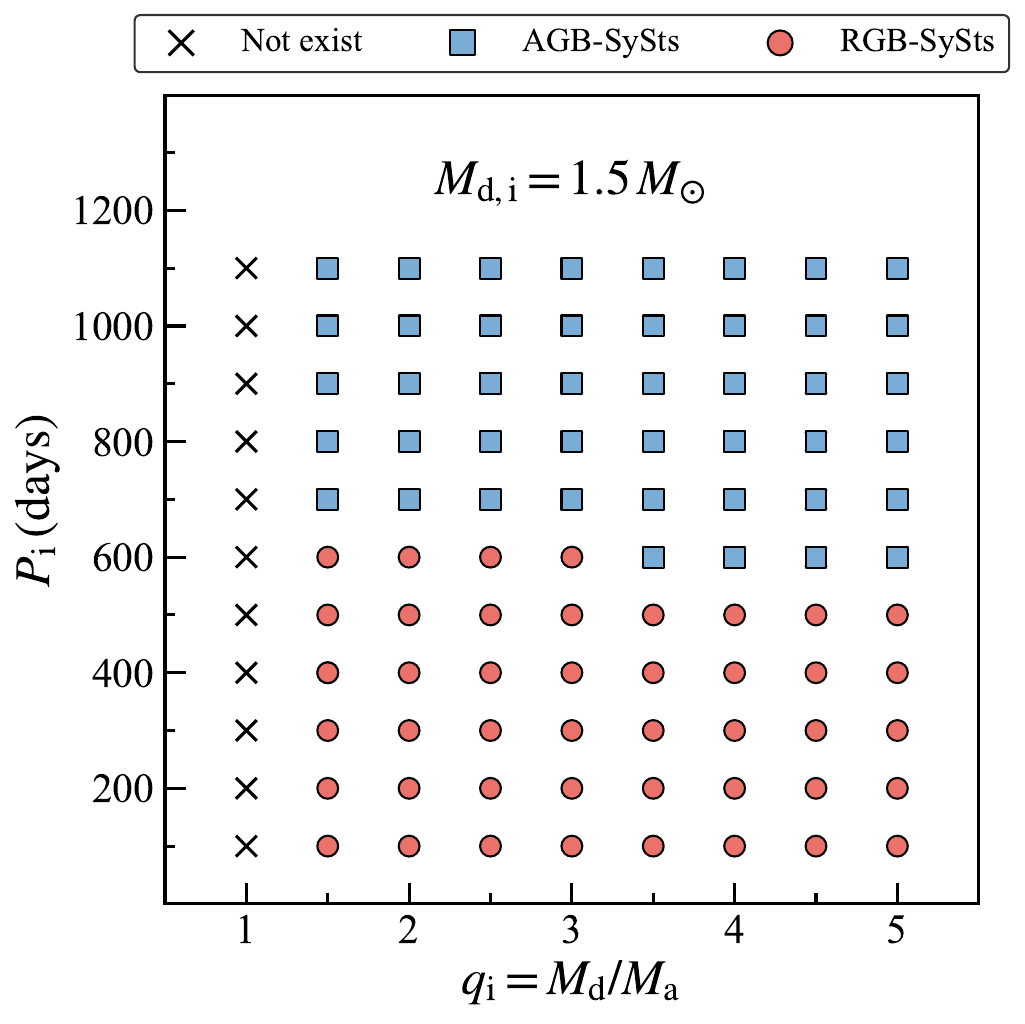}
        \label{fig:mass_1.5}
    \end{minipage}

    \vspace{0.5cm}

    \begin{minipage}{0.45\linewidth}
        \centering
        \includegraphics[width=0.9\linewidth]{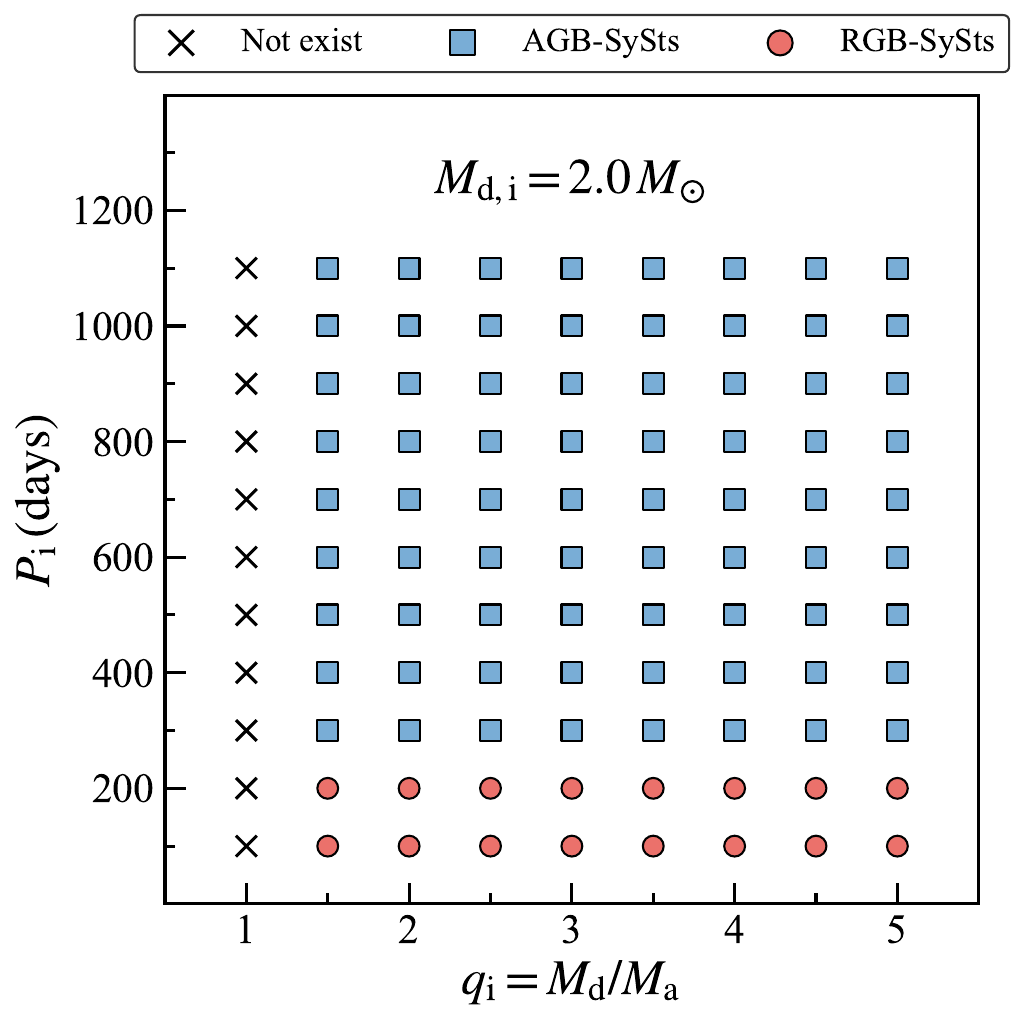}
        \label{fig:mass_2.0}
    \end{minipage}
    \begin{minipage}{0.45\linewidth}
        \centering
        \includegraphics[width=0.9\linewidth]{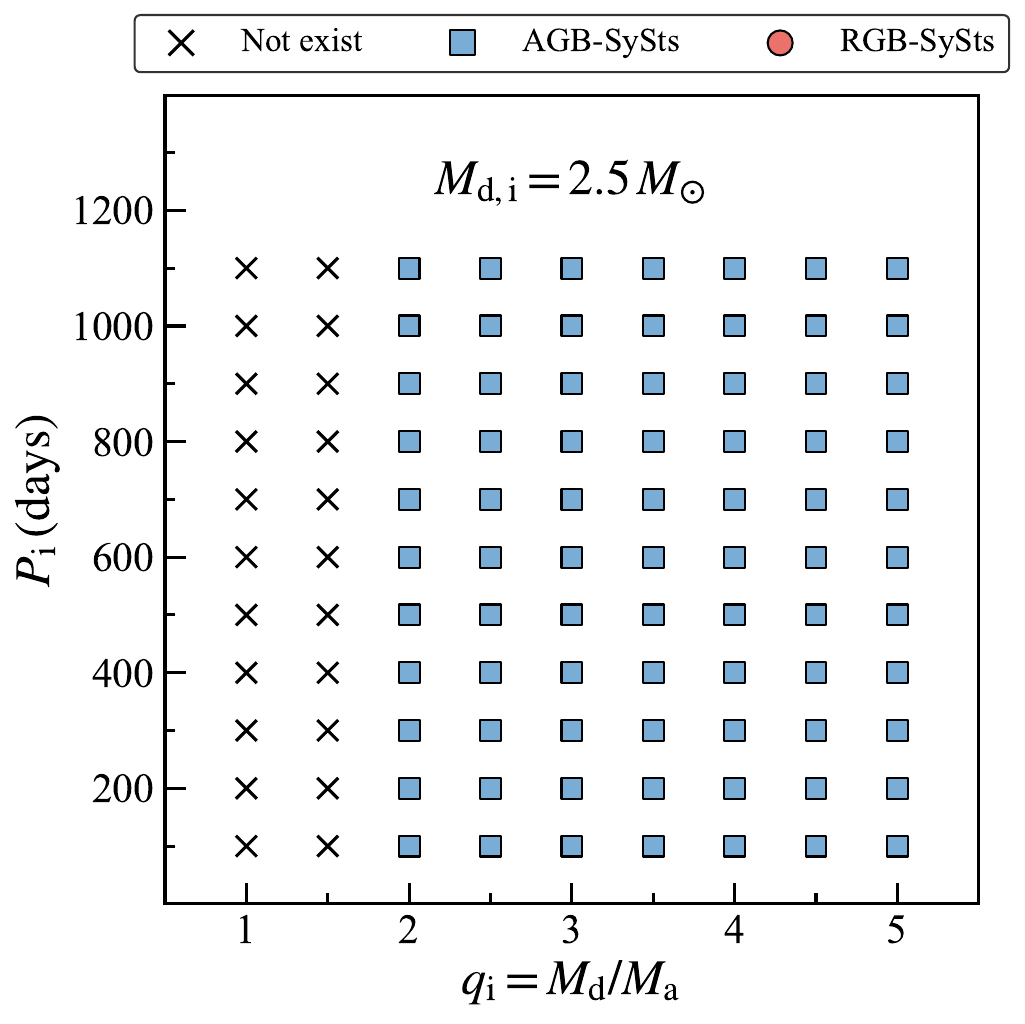}
        \label{fig:mass_2.5}
    \end{minipage}

    \vspace{0.5cm}

    \begin{minipage}{0.45\linewidth}
        \centering
        \includegraphics[width=0.9\linewidth]{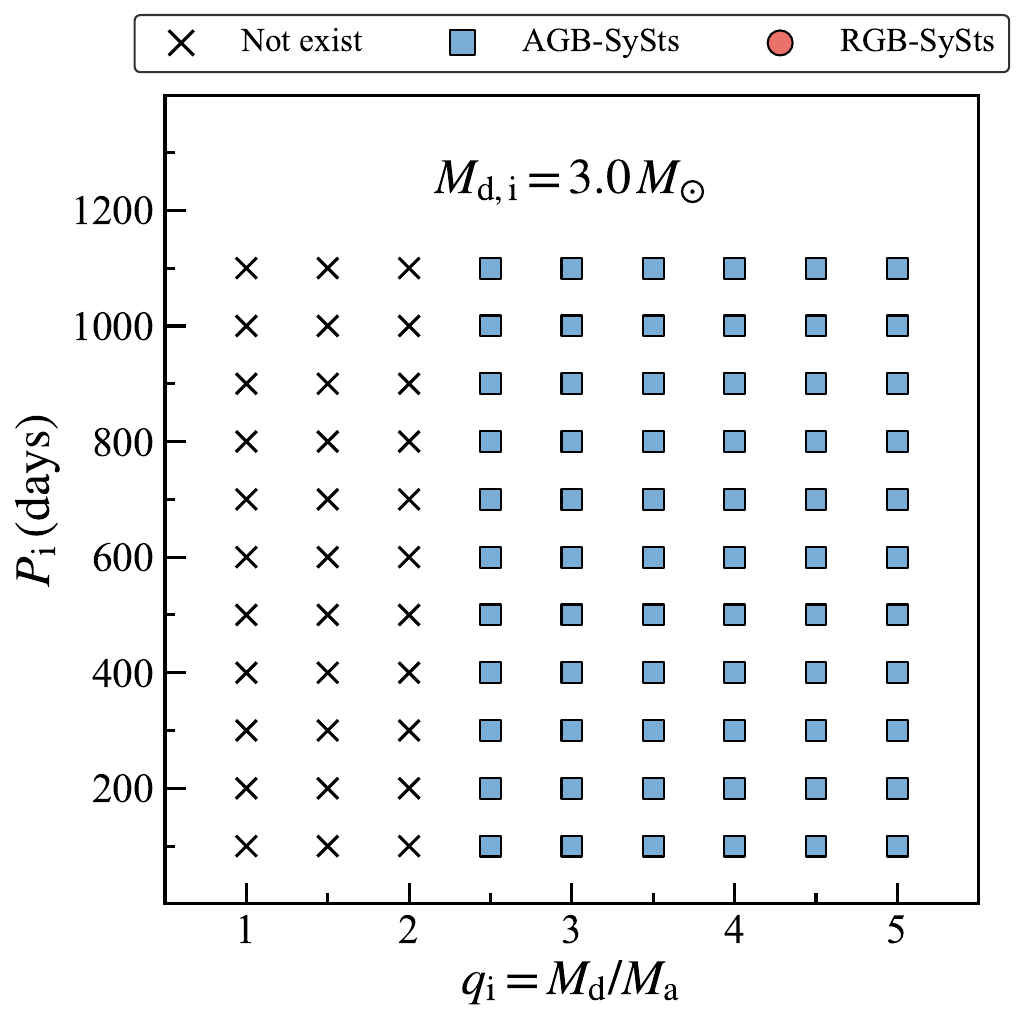}
        \label{fig:mass_3.0}
    \end{minipage}

    \caption{Binary grid showing the system types. The horizontal axis shows the initial mass ratio $q_{\rm i}$, while the vertical axis shows the initial orbital period $P_{\rm i}$. Crosses mark systems that do not become SySts because the initial accretor mass exceeds the Chandrasekhar limit. Red circles denote SySts with RGB donors, while blue squares denote SySts with AGB donors. Different panels correspond to models with different initial donor masses.}
    \label{3}
\end{figure*}

\noindent As described in Section~\ref{sec:binary grid parameter}, we constructed a grid of binary models covering a range of donor masses, mass ratios, and orbital periods. Figure~\ref{3} summarizes the results  and shows  how symbiotic systems can be classified by the evolutionary stage of the donor star at which they enter the symbiotic phase, which depends sensitively on these initial parameters. In this study, we define the onset of the symbiotic phase as the moment when the mass transfer rate exceeds $ 10^{-9}\,M_{\odot}\,\text{yr}^{-1}$, since only then can emission lines be observed against the strong continuum of the giant.

\par The evolutionary stage at which a system enters the symbiotic phase is determined primarily by the initial binary parameters, particularly the donor mass and the orbital period. Low-mass donors typically fill or approach their Roche lobes while ascending the RGB, allowing mass transfer rates high enough to produce observable symbiotic activity. For intermediate-mass donors, the outcome depends on the orbital period: in close binaries, the donor’s radius becomes comparable to its Roche-lobe radius already during the RGB, enabling an early onset of mass transfer; in wider systems, the donor remains well within its Roche lobe throughout the RGB and therefore does not reach the symbiotic phase until the asymptotic giant branch, when its radius expansion finally leads to significant mass loss. More massive donors usually remain detached on the RGB and only approach their Roche lobes on the asymptotic giant branch, where thermal pulsations facilitate mass transfer. Consequently, the diversity of donor types in our model grid reflects how the initial orbital separation and donor mass jointly determine whether and when the star approaches its Roche lobe, and thus the onset and duration of the symbiotic phase.

\par
\begin{figure*}[htbp]
    \centering
    \begin{minipage}{0.4\linewidth}
        \centering
        \includegraphics[width=0.9\linewidth]{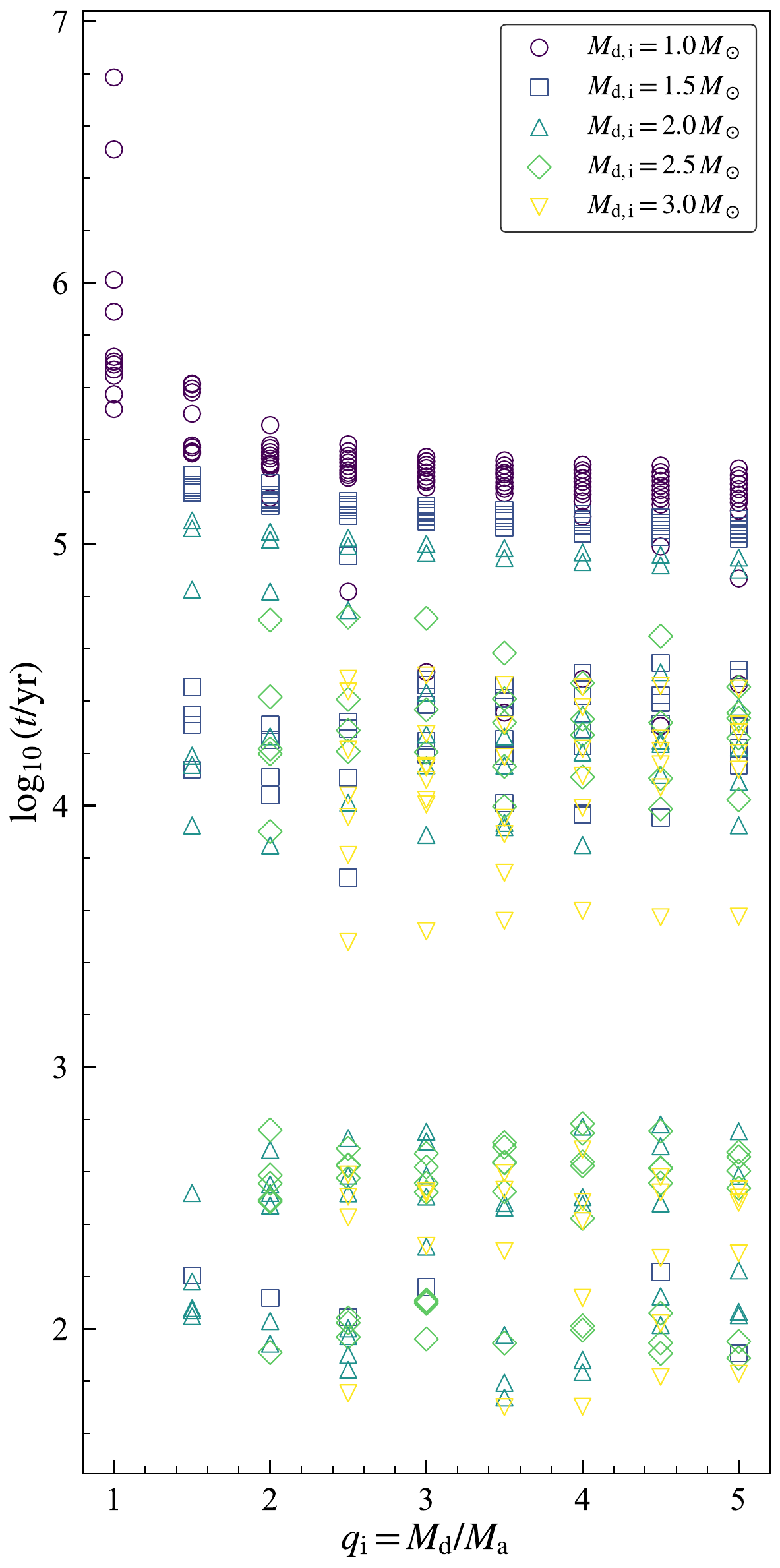}
    \end{minipage}
    \begin{minipage}{0.4\linewidth}
        \centering
        \includegraphics[width=0.9\linewidth]{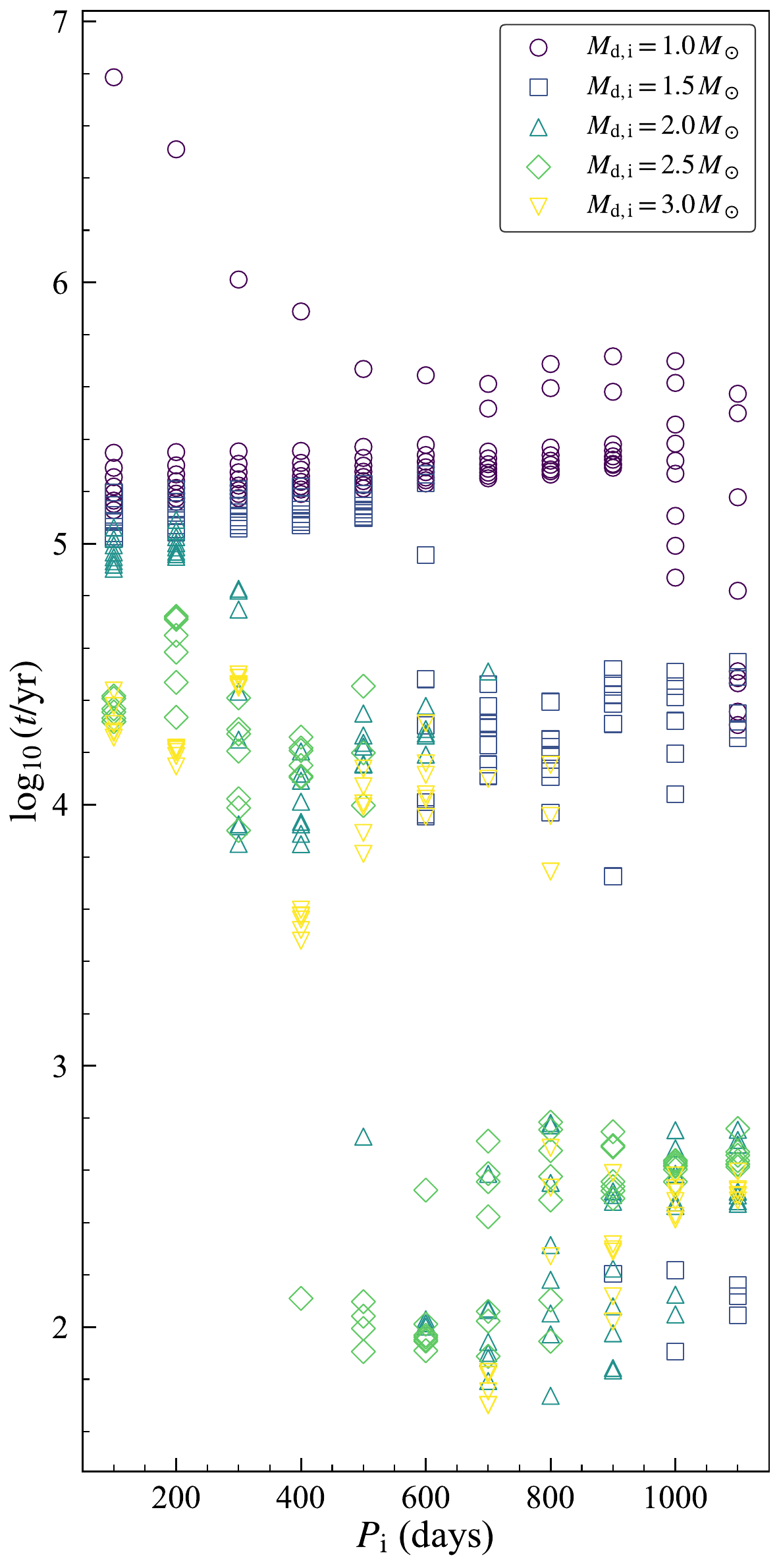}
    \end{minipage}
    \caption{Timescale variation with initial mass ratio $q_\mathrm{i}$ and initial orbital period $P_\mathrm{i}$. Different colors and symbols represent systems with different initial donor masses $M_\mathrm{d,i}$. For systems with AGB donors, the total timescale is the sum of individual thermal pulse periods. The left panel shows the timescale as a function of $q_\mathrm{i}$, while the right panel shows the timescale as a function of $P_\mathrm{i}$.}
    
    \label{4}
\end{figure*}

 Figure~\ref{4} shows the SySt lifetime of different models with various initial orbital periods ($P_\mathrm{i}$) and initial mass ratios ($q_\mathrm{i} = M_{\rm d}/M_{\rm a}$).

\par To quantify these trends, we examined how the duration of the symbiotic phase depends on the initial mass ratio and orbital period across the entire model grid. For systems with low mass ratios ($q \lesssim 1.5$), the symbiotic-phase lifetime typically lies in the range $1.37\times10^5$--$2.35 \times 10^6$~yr, with the longest-lived models reaching 
$6.1 \times 10^6$~yr. 
These long durations occur because the mass transfer rate settles into a long-lived, self-regulated state. Once the mass ratio, defined here as the donor mass divided by the accretor mass, drops below unity, the donor first becomes less massive than the accretor. This mass-ratio reversal causes the orbit to begin to re-expand, which further stabilizes the mass transfer. The dependence on the initial orbital period within this low-$q$ regime is relatively weak, provided that the donor fills (or nearly fills) its Roche lobe on the RGB.

\par At intermediate and high mass ratios ($q \approx 2$--4), the behaviour changes qualitatively. The onset of mass transfer causes rapid orbital shrinkage, which in turn drives a steep rise in the mass transfer rate and pushes the system toward $L_2$ overflow on a much shorter timescale. Across this regime, the symbiotic phase typically lasts 
$9.27\times10^3-2.26\times 10^5$~yr, with representative values around 
$1.53\times10^5$~yr for systems with $q \sim 2.5-3$. Only a small subset of short-period binaries reaches durations up to $2.51\times10^5$~yr before entering the dynamical regime. Even these shorter lifetimes are one to two orders of magnitude longer than the $\sim 10^3$~yr 'pre-CE' durations commonly assumed in population-synthesis studies.

\par We find that for systems with initial donor mass $M_\mathrm{d,i}=1.0\, M_\odot$ the timescale generally decreases as $q_\mathrm{i}$ becomes larger. For systems with larger initial mass ratios, the onset of mass transfer leads to faster orbital shrinkage. This accelerates the evolution through the SySt phase, so the total timescale becomes shorter. The timescales are the largest for mass ratios below $q_\mathrm{i}=1.5$, because most of these systems enter a stable mass transfer phase and sustain a high mass transfer rate for a longer time.

\par Systems with AGB donors behave differently because the mass transfer rate exceeds the observational threshold only during thermal pulses. For such binaries, the total observable lifetime is the sum of several short episodes, each lasting about $100$~yr, separated by interpulse phases during which the donor does not approach its Roche lobe. 
During these pulses, the mass transfer rate can exceed $10^{-9} \, M_{\odot} \, \text{yr}^{-1}$, and can also fall below this threshold (see Section~\ref{The timescale of AGB-SySts}). Because of this variability, the total timescale of these systems is a sum over all individual thermal pulse periods. 
For AGB donors, neglecting stellar winds has two competing effects. On one hand, wind mass loss can shorten the available AGB lifetime by removing the envelope. On the other hand, non-conservative mass and angular momentum loss can slow the orbital shrinkage and delay the onset of rapid mass transfer, allowing the binary to remain longer in the adopted symbiotic mass transfer rate range. Therefore, the AGB evolutionary lifetimes and the SySt residence times should not be interpreted in the same way; the former may be overestimated, while the latter are likely lower limits within our conservative orbital-evolution treatment.

\par Although multiple SySt episodes can occur, their total duration remains short. This effect, namely that a system does not appear as a SySt continuously during the advanced evolutionary stage, is evident in  Figure~\ref{4}, where the symbiotic phase timescale drops sharply for systems with $M_\mathrm{d,i}=1.0\,M_\odot$ once  $P_\mathrm{i}>1100\,\mathrm{days}$; for $M_\mathrm{d,i}=1.5\,M_\odot$ once $P_\mathrm{i}>600\,\mathrm{days}$; and for $M_\mathrm{d,i}=2.0\,M_\odot$ once $P_\mathrm{i}>200\,\mathrm{days}$.

\subsection{Orbital period change}

\begin{figure*}[htbp]
\centering
\includegraphics[scale=0.35]{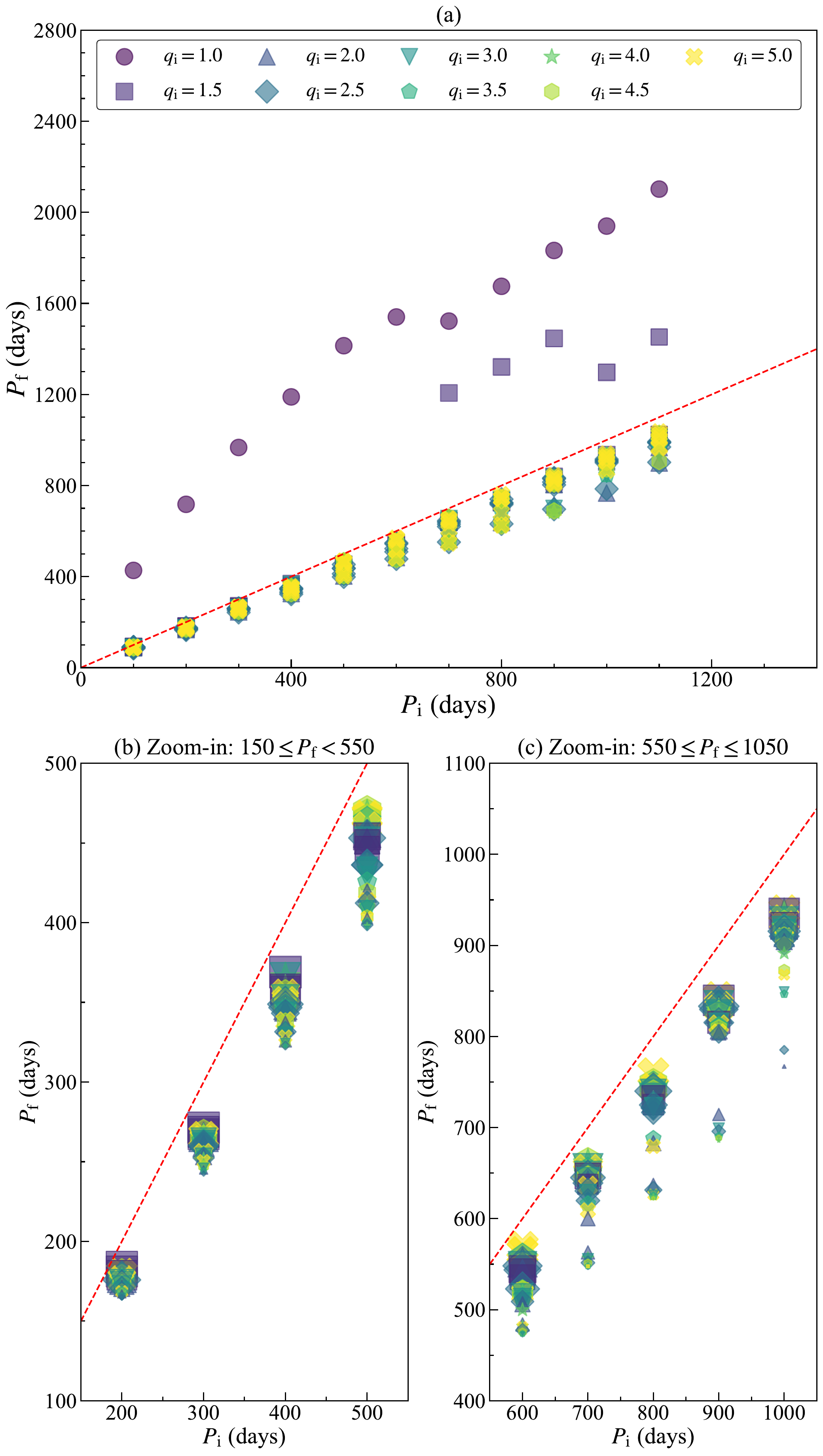}
\caption{Changes in the orbital period. Panel (a) shows all systems in the $P_\mathrm{f}$-$P_\mathrm{i}$ diagram. The red dashed line represents systems where the final orbital period $P_\mathrm{f}$ equals the initial orbital period $P_\mathrm{i}$. Panels (b) and (c) present enlarged views of the short- and long-period regions, respectively. Different colors and symbols denote models with different initial mass ratios $q_\mathrm{i}$. The symbol sizes in panels (b) and (c) have no physical meaning and are adjusted only to improve visual clarity.}
\label{5}
\end{figure*}

\noindent Since the orbital period and binary separation change during mass transfer, it is important to discuss how the orbital period evolves in order to better understand its connection to the mass transfer timescale. In Figure~\ref{5}, we show the changes from the initial to the final orbital periods.
We note that these trends reflect the assumption of fully conservative mass transfer; in reality, mass and angular momentum loss through winds, disc outflows, or nova eruptions could alter the orbital evolution.
As expected for conservative mass transfer, in most systems with $q_\mathrm{i}\ge 2$, $P_\mathrm{f}$ decreases to about $0.8$ of $P_\mathrm{i}$. In contrast, most systems with $q_\mathrm{i}= 1.5$, and some with $q_\mathrm{i}\le1.5$, experience an increase in their orbital periods.
The orbital period increases because mass transfer from the donor to the accretor reduces the mass ratio below 1.0. This causes the binary separation to grow and the orbital period to lengthen \citep{tauris2023physics}. The combined effect of donor expansion and orbital widening leads to a stable mass transfer phase that lasts for a long time.

\subsection{Comparison with observation}

\noindent To compare with theoretical models, we selected a sample of 45 Galactic SySts containing accreting WDs, as shown in Table \ref{tab:observation}. All systems in this sample have well-determined orbital periods, estimated component masses, identified mass transfer modes (Roche lobe overflow or atmospheric Roche lobe overflow), and inferred mass accretion rates. Systems
with neutron star accretors, extragalactic SySts, or missing key parameters were excluded.

\par Among the 45 systems, 36 have orbital periods shorter than $\sim$1000 days. At least 20 (and possibly two more) of them show ellipsoidal light curves (ELC), which suggest that their red giant donors are filling or nearly filling ($\gtrsim 80\%$) their Roche lobes. Detection of ellipsoidal LCs requires orbital inclinations $\gtrsim 60^\circ$, so additional systems with low inclinations and small mass functions $f(m_g)$ may also be close to Roche lobe filling but lack detectable ellipsoidal variations.

\par This fraction of ellipsoidal systems agrees well with the statistics from the broader population of SySts with accreting WDs and known photometric periods. Currently, 149 such systems are known; 113 of them have $P \lesssim 1000$ days, and 64 show ELC (Ilkiewicz et al., in preparation).

\par The estimated masses of red giant donors in our sample range from $\sim$1 to 2.5~$M_\odot$, while the mass ratios ($q = M_\mathrm{d}/M_\mathrm{a}$) typically range between $\sim$1.5 and 4. An exception is the group of symbiotic recurrent novae, which have $q < 0.8$; three such systems (T~CrB, RS~Oph, V3890~Sgr) have orbital periods below 1000 days. The WD accretion rates, derived from luminosities, are all higher than $\sim10^{-8}\ M_\odot\,\mathrm{yr}^{-1}$, and many reach $\sim10^{-7}\ M_\odot\,\mathrm{yr}^{-1}$. These three systems all belong to recurrent novae, in which a white dwarf experiences repeated thermonuclear runaways on its surface due to continued accretion from their companion star. As shown in Table~\ref{tab:observation}, all three systems have giant masses $M_\mathrm{g}$ lower than the white dwarf masses $M_\mathrm{WD}$. In Figure~\ref{fig:9}, we mark these systems on the $P-q$ diagram and select the models whose evolutionary tracks closely match the observations. We find that in these models, the final accretor mass $M_\mathrm{a,f}$ is larger than the final donor mass $M_\mathrm{d,f}$. RS~Oph matches remarkably well, with modeled $M_\mathrm{d}$ and $M_\mathrm{a}$ almost identical to the observed values. T~CrB shows a slight difference between modeled and observed masses, but the agreement is still close. V3890~Sgr exhibits a larger discrepancy, with both $M_\mathrm{d}$ and $M_\mathrm{a}$ in our models being lower than the observed values.

\begin{figure*}[htbp]
    \centering

    \begin{subfigure}{0.30\linewidth}
        \centering
        \includegraphics[width=\linewidth]{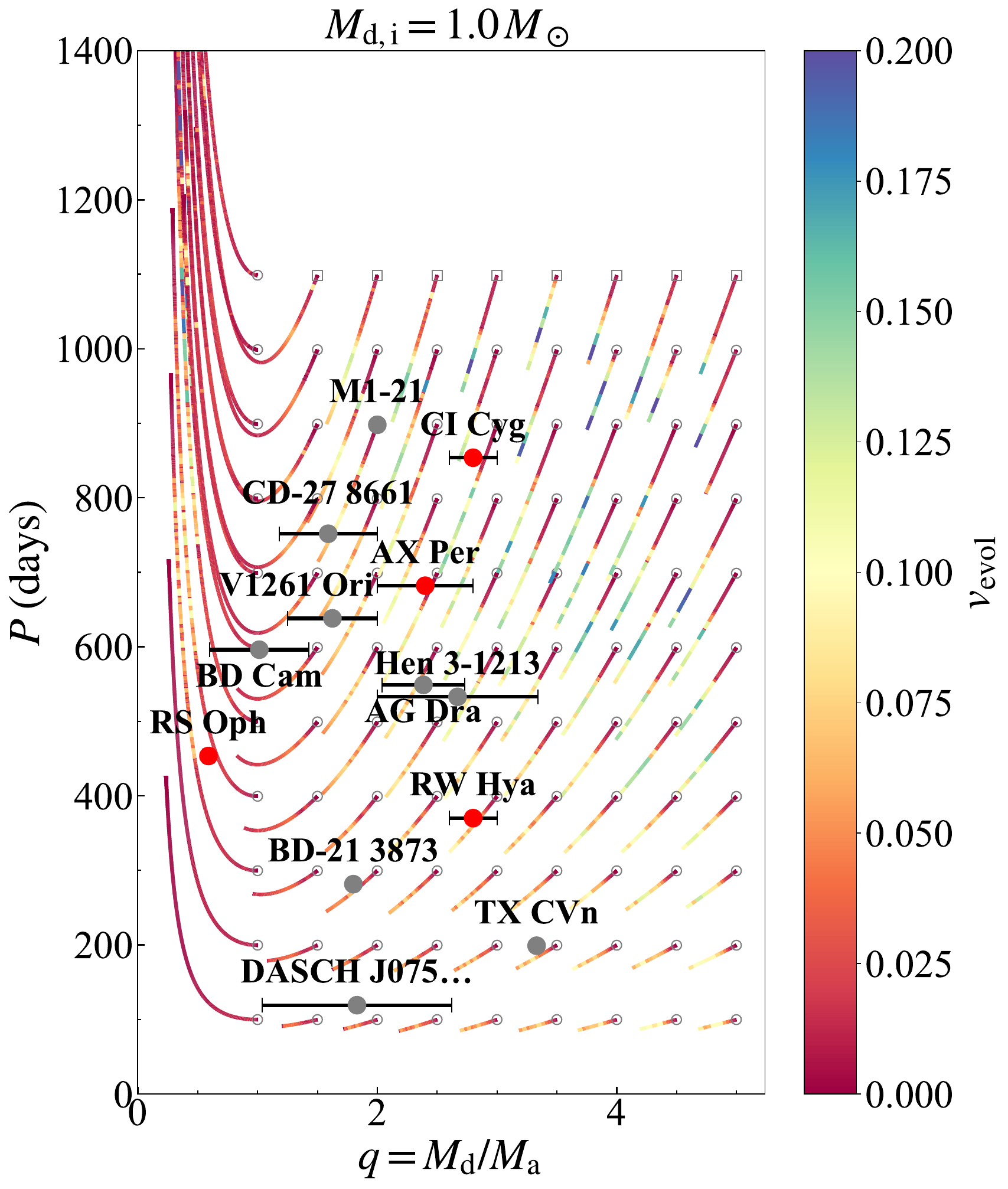}
    \end{subfigure}
    \hfill
    \begin{subfigure}{0.30\linewidth}
        \centering
        \includegraphics[width=\linewidth]{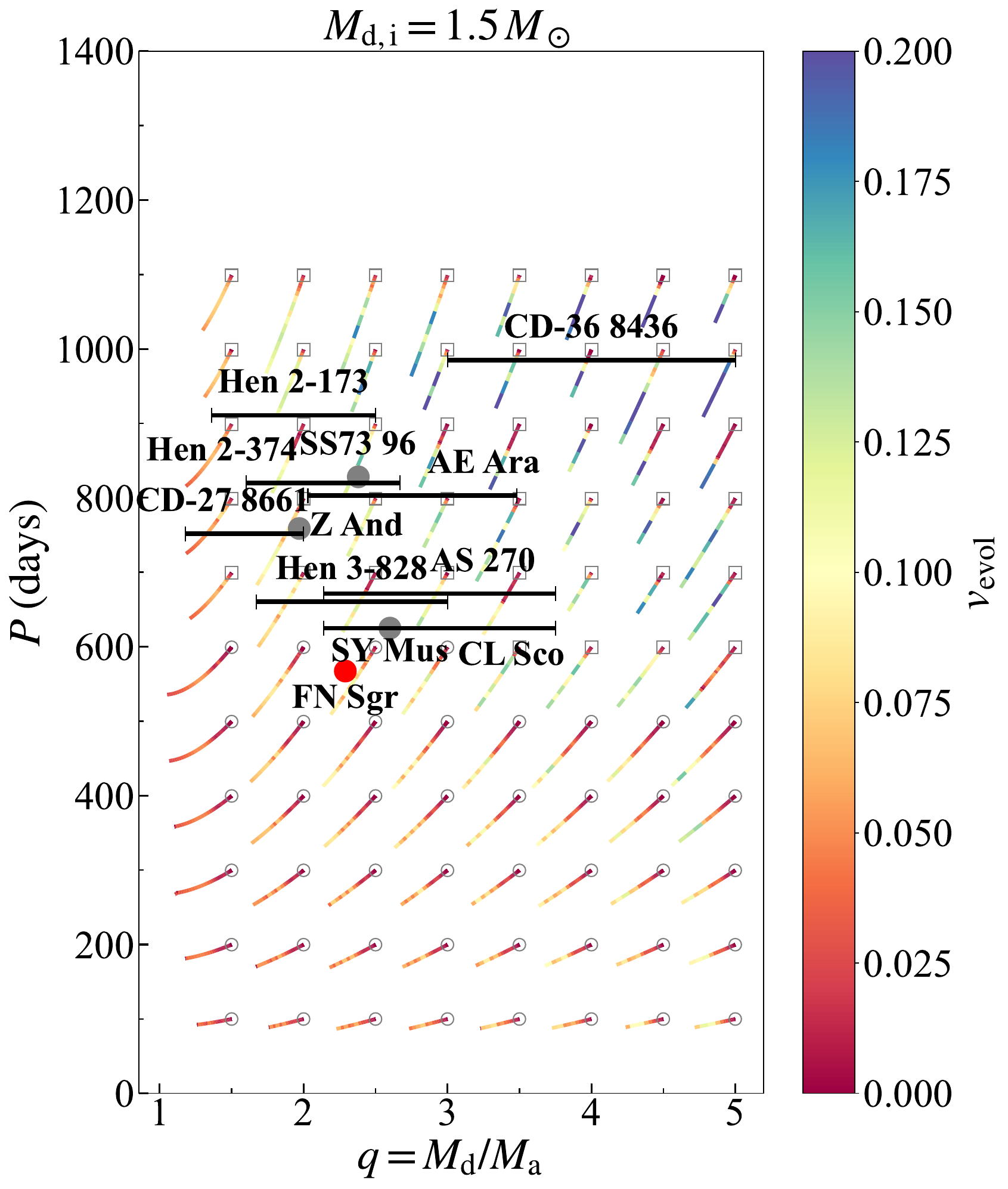}
    \end{subfigure}
    \hfill
    \begin{subfigure}{0.30\linewidth}
        \centering
        \includegraphics[width=\linewidth]{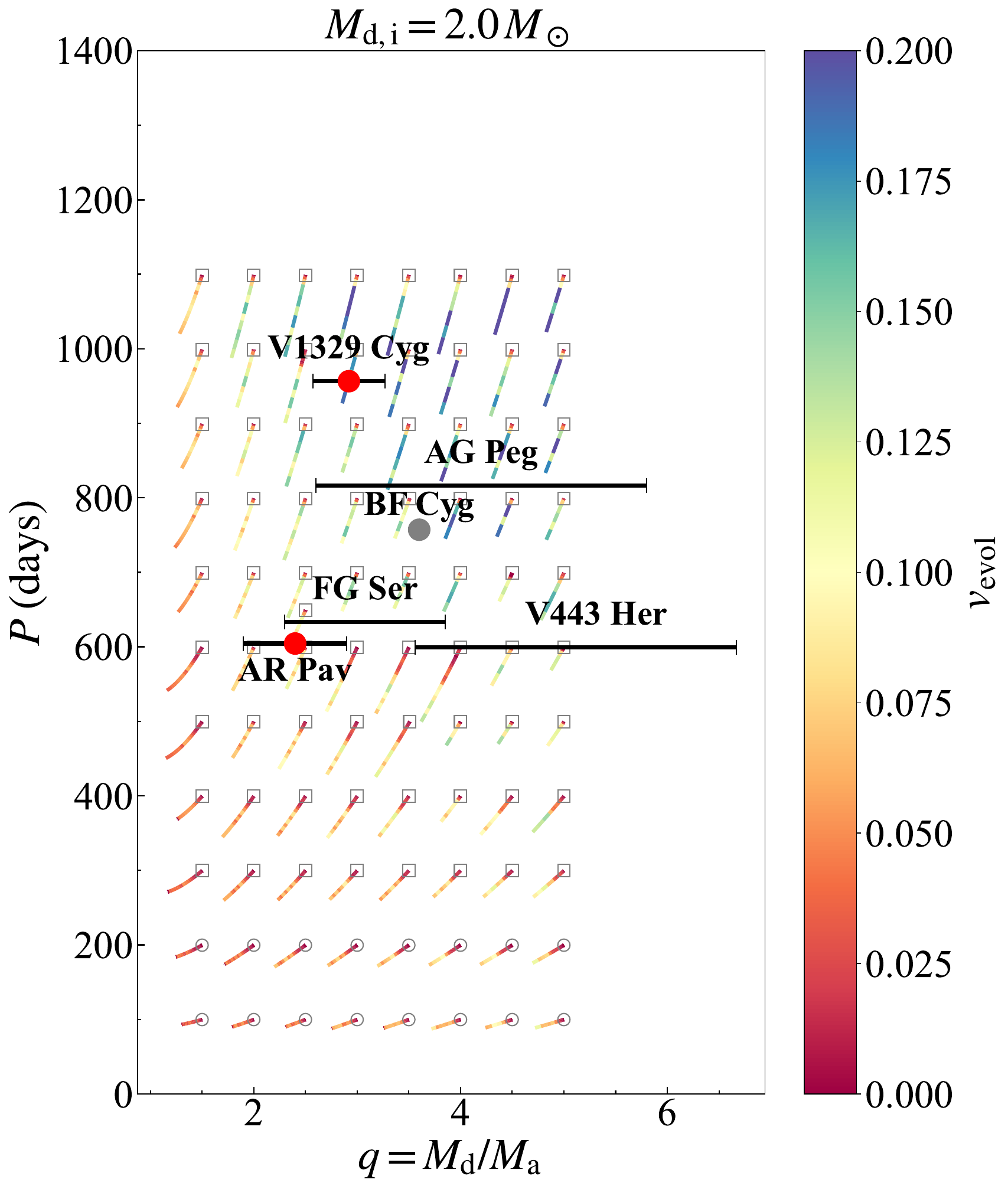}
    \end{subfigure}

    \vspace{0.35cm}

    \begin{subfigure}{0.30\linewidth}
        \centering
        \includegraphics[width=\linewidth]{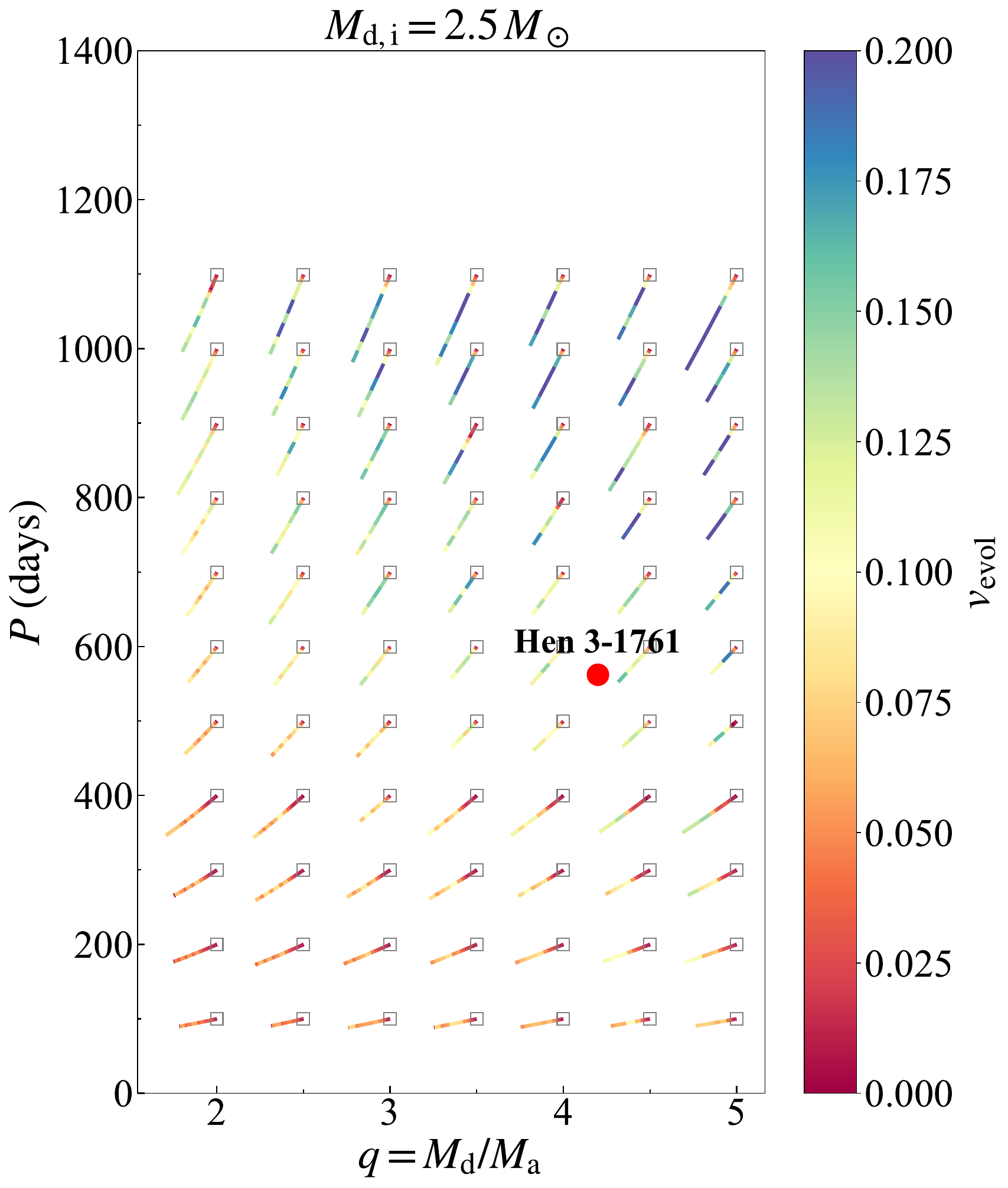}
    \end{subfigure}
    \hspace{0.08\linewidth}
    \begin{subfigure}{0.30\linewidth}
        \centering
        \includegraphics[width=\linewidth]{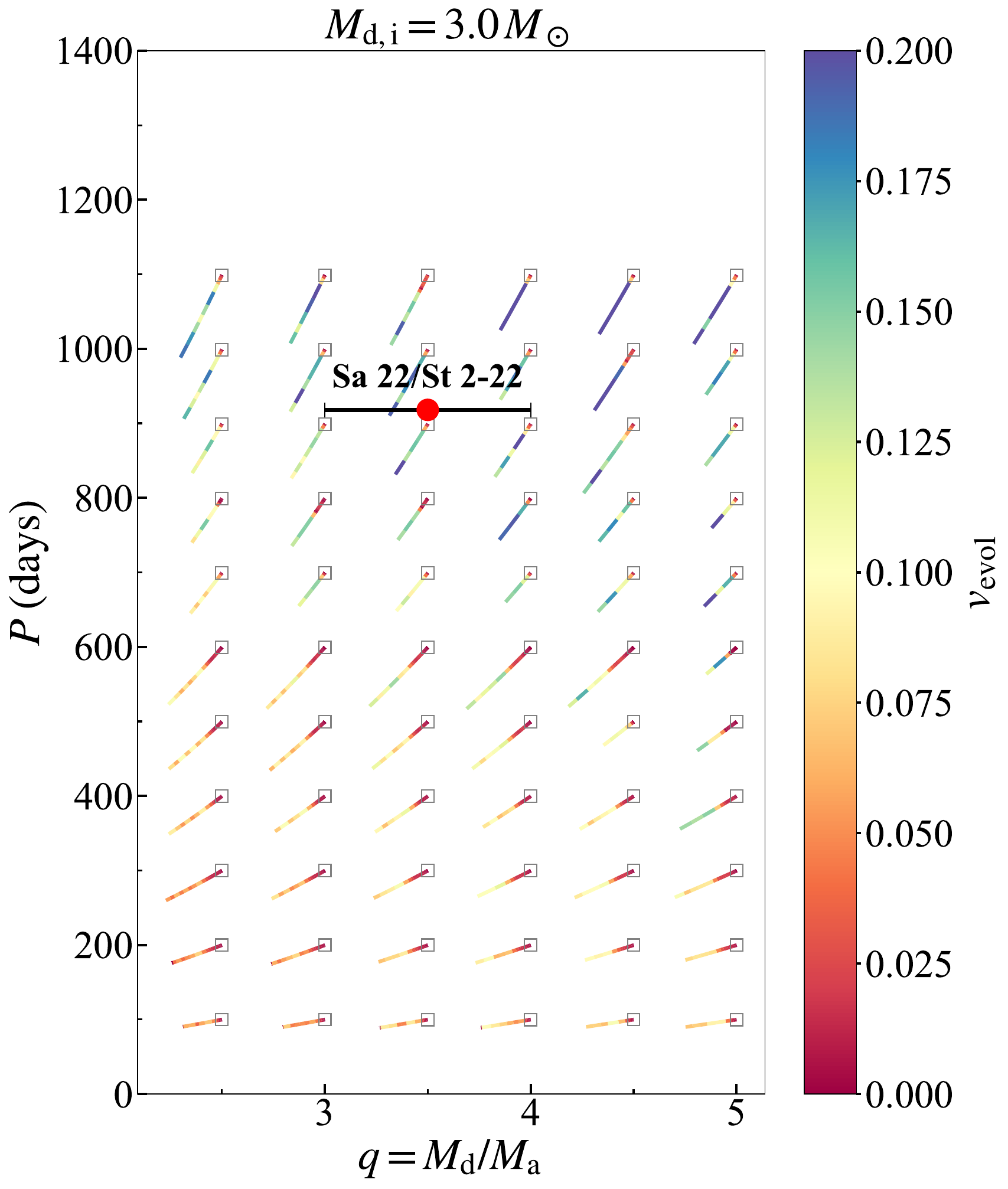}
    \end{subfigure}

    \caption{Comparison between our model predictions and observational data in the $P-q$ plane. Gray circles and squares denote model systems entering the SySt phase during the RGB and AGB phases, respectively. The color scale shows the diagram-based evolutionary speed, $v_{\rm evo}$, defined in the text. This quantity is interpreted as an inverse residence time: models with lower $v_{\rm evo}$ move more slowly through the $P-q$ plane and therefore spend longer in the corresponding region. Red filled circles show observed systems with uncertainties.}
    \label{fig:6}
\end{figure*}

Figure~\ref{fig:6} compares our model predictions with the observed systems described above. To interpret their distribution in the $P-q$ plane, we quantify the evolutionary speed of each model across this diagram, using the temporal changes in orbital period and mass ratio. This speed is best interpreted as an inverse residence time: regions where systems move slowly correspond to evolutionary phases in which binaries spend more time, and are therefore more likely to be observed, whereas rapidly crossed regions are less likely to be populated observationally.

For two consecutive outputs at times  \(t_i\) and \(t_{i+1}\), we evaluate the changes in the plotted quantities,
\begin{equation}
\Delta P = P_{i+1} - P_i, \qquad
\Delta q = q_{i+1} - q_i, \qquad
\Delta t = t_{i+1} - t_i .
\end{equation}
To ensure that both axes contribute comparably to the measured evolutionary rate, we scale the changes by the characteristic mesh size used in the plot, adopting
\begin{equation}
\Delta P_m = 100~\mathrm{days}, \qquad
\Delta q_m = 0.5 .
\end{equation}
We then define a diagram-based evolutionary speed as
\begin{equation}
v_{\rm evo}
= \frac{1}{\Delta t}
  \sqrt{
        \left( \frac{\Delta P}{\Delta P_m} \right)^{2}
        +
        \left( \frac{\Delta q}{\Delta q_m} \right)^{2}
       } .
\end{equation}

This quantity is a diagnostic of how rapidly the evolutionary track changes its
position in the $P$–$q$ plane, providing a convenient measure of the local rate
of evolutionary change in the plotted coordinates.
We select systems in which the giant mass ($M_{\rm g}$) lies within approximately $0.3\,M_\odot$ of the initial $M_{\rm g}$ in our model. Additionally, we only include observed systems with orbital periods shorter than 1100 days, as these are more likely to be S-type SySts.  From Figure~\ref{fig:6}, we can see that our parameter space could cover most of the observational points.

Note that our models assume fully conservative mass transfer. Non-conservative mass transfer may still allow systems to occupy similar regions of the $P-q$ plane, but it can change how rapidly they evolve through this plane. Thus, the comparison with observed systems should be interpreted not only in terms of track location, but also in terms of residence time in different regions of the diagram. We also emphasize that coverage of the $P-q$ plane alone is insufficient to determine whether our models can reproduce all of the observed symbiotic systems. The current component masses must also be matched. Taken together, the orbital period, mass ratio, and component masses provide stronger constraints on the possible progenitor systems of the observed symbiotics.

\begin{figure*}[t]
  \sidecaption
  \resizebox{12cm}{!}{
    \begin{tabular}{cc}
      \includegraphics[width=0.48\linewidth]{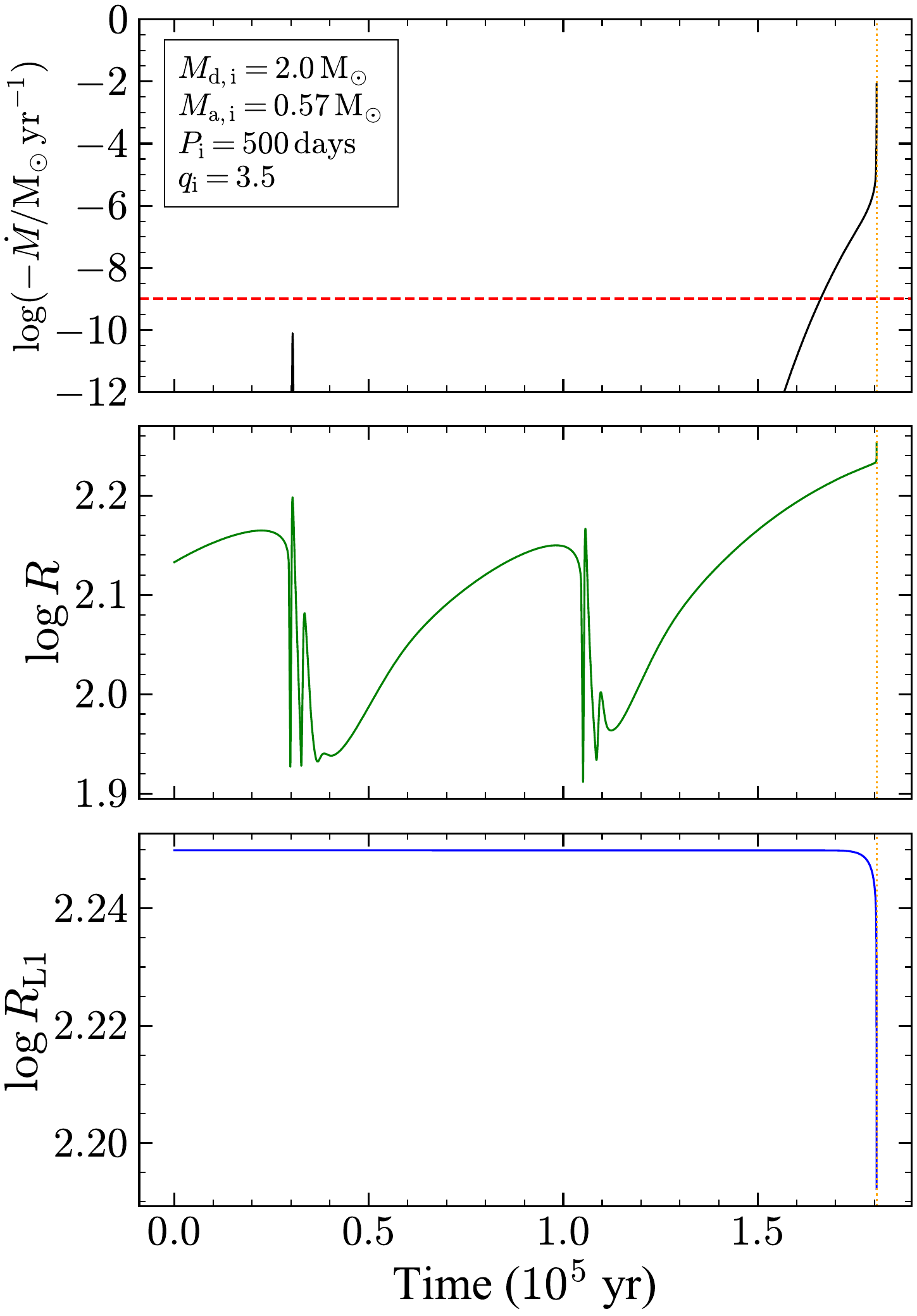} &
      \includegraphics[width=0.49\linewidth]{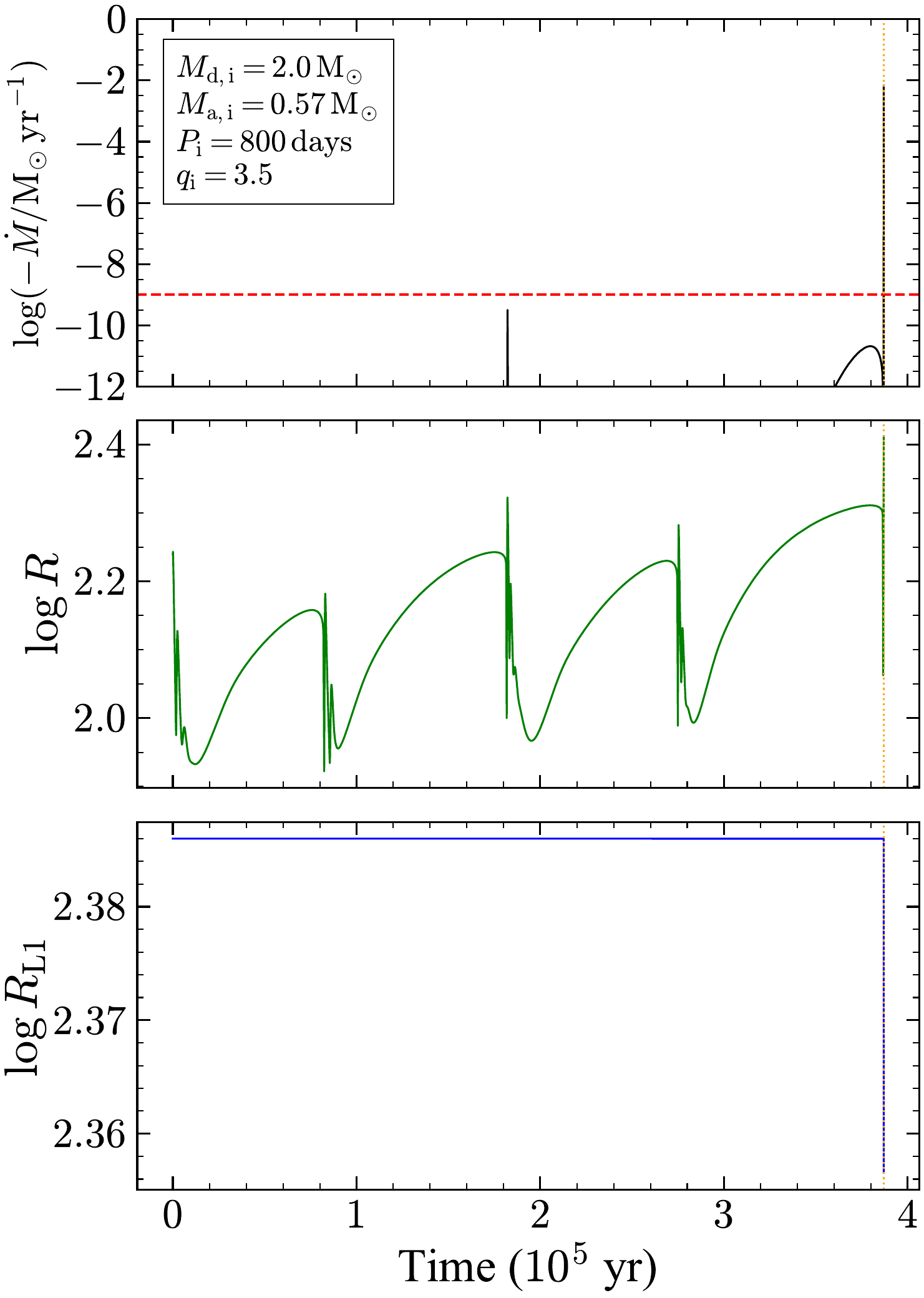}
    \end{tabular}
  }
  \caption{Examples of AGB-SySts with different mass-transfer timescales. The left and right panels show binaries initially composed of a $2.0\,M_{\odot}$ giant and a $0.57\,M_{\odot}$ WD, with initial orbital periods of 500 and 800 days, respectively. The red dashed line marks $\dot{M}=10^{-9}\,M_{\odot}\,\mathrm{yr}^{-1}$, while the orange dotted line indicates when the system exceeds the $L_2$ equipotential radius and stable mass transfer stops.}
  \label{fig:7}
\end{figure*}

\par We have selected five observationally well-determined symbiotic systems, CI Cyg, FN Sgr, AR Pav, V1329 Cyg, and RW Hya (see Table \ref{tab:observation}), to compare with our theoretical evolutionary tracks, as shown in Figure~\ref{8}. All five systems exhibit eclipsing binary characteristics, which greatly reduce the uncertainties in their mass determinations. The comparison shows excellent agreement between the observed systems and our model evolutionary tracks, satisfying the criteria of orbital period within 50 days, donor mass within 0.25 $M_\odot$, and mass ratio within 0.5 of the modeled values along the track. These results reinforce the predictive strength of our models for the evolutionary pathways of symbiotic binaries.

\section{Discussion} \label{Discussion}
\label{sec:discussion}

\subsection{The timescale of AGB-SySts}\label{The timescale of AGB-SySts}

\noindent Figure~\ref{fig:7} shows that for systems with shorter orbital periods, the donor star undergoes thermal pulses which cause its radius to vary. During the first few thermal pulses, the star expands and the mass transfer rate reaches $10^{-9}\,M_{\odot}\,\text{yr}^{-1}$ at the end of each cycle. During the final, stronger thermal pulse, the star expands more rapidly, allowing the mass transfer rate to exceed $10^{-9}\,M_{\odot}\,\text{yr}^{-1}$ earlier in the cycle, resulting in a longer mass transfer timescale. In contrast, for systems with longer initial periods, even strong thermal pulses cannot bring the donor's radius close enough to fill its Roche lobe, so the mass transfer timescale remains short.

In this study, we did not include the effects of stellar winds, disc outflows from the white dwarf, or nova-driven mass and angular-momentum loss on the orbital evolution. This approximation is reasonable for RGB donors, but becomes less accurate for AGB donors, for which substantial wind mass loss is expected. As shown in Figure~\ref{3}, some systems contain AGB donors and show variability in the mass-transfer rate associated with thermal pulses. In our models, the total duration of the AGB contribution to the SySt phase is obtained by summing the durations of the individual thermal-pulse episodes during which the mass-transfer rate exceeds
$10^{-9}\,M_\odot\,\mathrm{yr}^{-1}$.

Non-conservative mass and angular-momentum loss can modify the orbital evolution and therefore the duration of individual mass-transfer episodes. Stellar winds can also shorten the TP-AGB lifetime by removing the envelope and reducing the total number of thermal pulses. However, in our calculations, AGB donors usually lose their envelopes through Roche-lobe overflow during the first few thermal-pulse episodes and rarely evolve beyond three pulses, whereas single AGB stars of comparable mass may experience many more pulses. Therefore, wind mass loss would mainly affect later pulses that are not reached in our models, and is unlikely to substantially decrease the SySt durations reported here. The timescales presented here should therefore be regarded as closer to lower limits for the time spent in the SySt regime.

\par Numerous studies have examined mass transfer processes in SySts. For example, \citet{podsiadlowski2007origin,mohamed2012mass,abate2013wind,chen2018wind,belloni2020absence,vathachira2025exploring,ilkiewicz2019wind} studied wind-Roche lobe overflow (WRLOF) in SySts, suggesting that WRLOF is a highly efficient mass transfer mechanism. Since our models include an AGB donor, the WRLOF is expected to operate in these systems as well.  However, the effect is not expected to have a substantial impact on most S-type SySts.

\subsection{Uncertainties of this work}
\noindent One important limitation of our models is that we assume fully conservative mass transfer and neglect mass loss. As emphasized by \citet{2024ApJ...971...64I}, the outcomes of mass transfer are highly sensitive to the adopted treatment of angular momentum loss, which we have not investigated in detail. Although the new formalism itself does not directly predict final outcomes, it consistently produces different mass loss rates for the same degree of Roche-lobe underflow compared to the KR prescription, particularly during the stage when the mass transfer rate increases exponentially. The quantitative results, however, still depend on the nature of the accretor and the specific channel of angular momentum loss assumed. In practice, different prescriptions for angular momentum loss, such as outflows through the outer Lagrangian points or varying degrees of conservative versus non-conservative transfer, could significantly alter the evolutionary outcome.

\par In our models, we assume circular orbits and neglect the influence of eccentricity on the mass transfer process. Observationally, most S-type symbiotic systems indeed have nearly circular orbits, with eccentricities close to zero. However, several studies suggest that even modest eccentricities (of order $10^{-2}$) can affect binary evolution and may further amplify eccentricity over time. Previous work has also investigated how eccentricity modifies the mass transfer process. For example, \citet{sepinsky2007interacting} showed that in eccentric binaries, mass transfer is inherently episodic rather than continuous, since Roche-lobe overflow can only occur near periastron. This phase-dependent behavior introduces strong variability in the mass transfer rate and invalidates the steady-state assumption commonly adopted for circular systems. Moreover, mass transfer can either reinforce or counteract tidal effects: while tides generally promote circularization, discrete periastron mass exchange may instead increase eccentricity, allowing binaries to maintain non-zero eccentricities over long timescales. For certain parameter ranges, this timescale is comparable to the overall mass transfer timescale. Finally, the burst-like episodes of angular momentum and mass exchange inherent to eccentric mass transfer have direct implications for stability, as they may promote unstable behavior and accelerate the onset of a common-envelope phase relative to predictions based on circular-orbit models.

\begin{figure*}[t]
  \sidecaption
  \includegraphics[width=12cm]{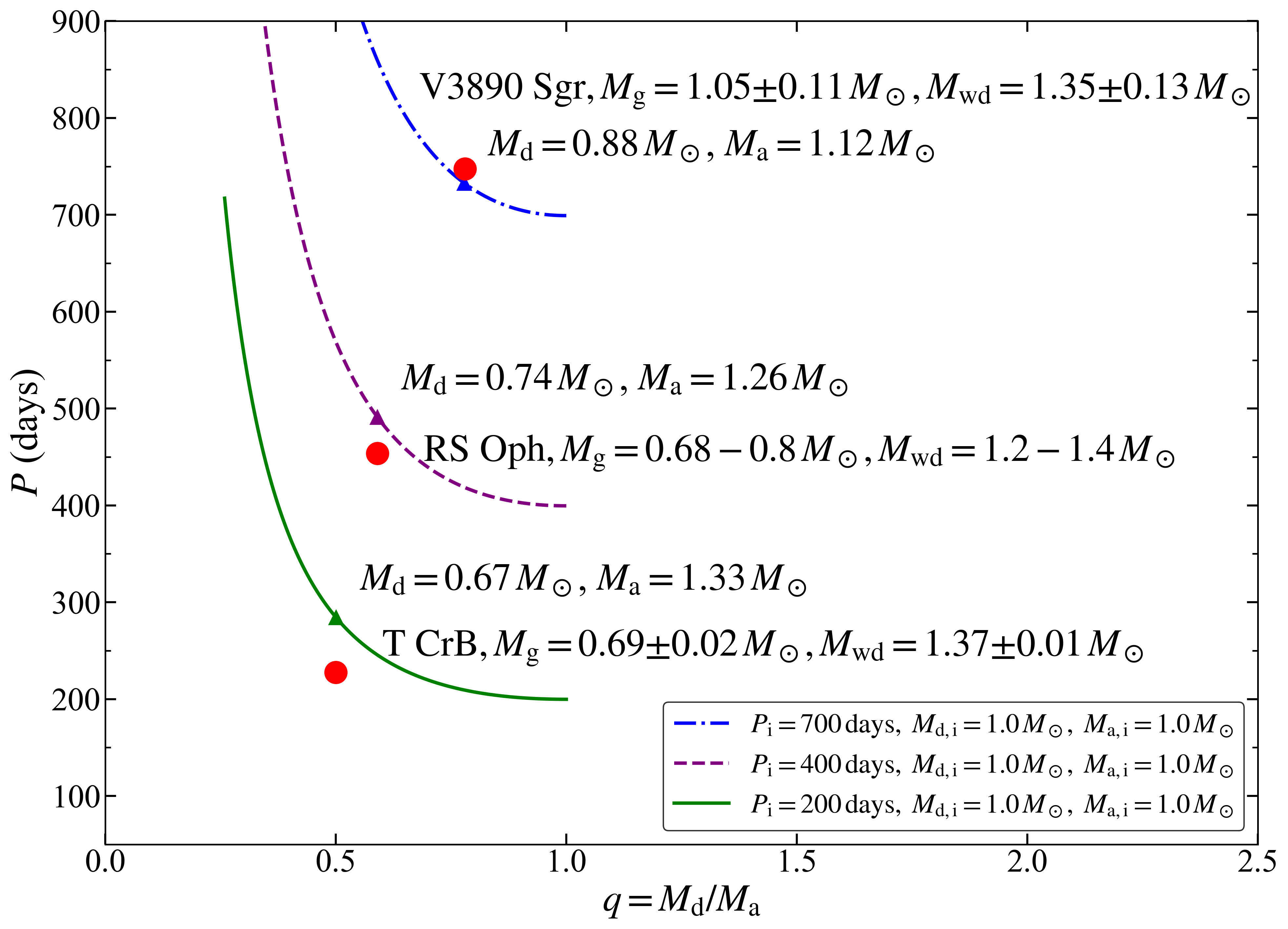}
  \caption{Positions of selected recurrent nova systems in the $P$--$q$ diagram, compared with our simulations. For each system, the corresponding mass-transfer track is shown, and the point where the mass ratio matches the observed value is marked with a triangle.}
  \label{fig:9}
\end{figure*}

\section{Conclusions}
\label{sec:conclusions}

\noindent This paper investigates mass transfer processes in SySts,  a class of interacting binaries composed of a giant donor and a compact accretor, typically a WD or a NS. Observational studies show that most donors in S-type SySts are nearly Roche lobe-filling, even in systems with long orbital periods, often exceeding several hundred days. However, standard binary evolution models predict that mass transfer in systems with high mass ratios ($q = M_{\mathrm{d}} / M_{\mathrm{a}} > q_{\mathrm{crit}}$) should be dynamically unstable and occur on short timescales, in contrast to what is observed. This persistent discrepancy between theoretical predictions and observations represents a long-standing challenge in understanding of SySts. RUMT avoids the inconsistencies in analytic atmospheric-overflow formalisms and provides physically robust mass transfer rates across both underflow and overflow regimes.

\par To address this issue, we employ an improved method for modeling mass transfer based on the RUMT framework, which unifies atmospheric RLOF with outflow through the $L_{\mathrm{1}}$ point. We use its implementation in the one-dimensional stellar evolution code \texttt{MESA} to perform simulations across a comprehensive grid of binary parameters, varying the initial donor mass $M_{\mathrm{d,i}}$ between 1 $M_{\odot}$ and 3 $M_{\odot}$, the initial mass ratio $q_\mathrm{i}$ ($M_{\rm d}/M_{\rm a}$) between 1 and 5, and the initial orbital period $P_\mathrm{i}$ between 100 and 1100 days. Our model assumes giant donors with extended atmospheres and examine how atmospheric RLOF can help sustain dynamically stable mass transfer.

\par The main results of this study are summarized as follows:

\begin{enumerate}

 \item The inclusion of atmospheric RLOF in the RUMT framework significantly extends the duration before the onset of dynamically unstable mass transfer, in some cases by up to three orders of magnitude compared with traditional theoretical models. The extended lifetimes found here remove the long-standing mismatch between the short ($\sim 10^3$~yr) pre-CE durations assumed in population studies and the observed number of S-type symbiotic systems.  These longer lifetimes naturally explain the high Galactic occurrence rate of S-type SySts and resolve the primary discrepancy in previous population-synthesis studies.

\item We find that the mass transfer timescale is sensitive to the initial conditions, particularly the mass ratio and orbital period:

\begin{itemize}
\item For systems with low mass ratios, mass transfer can remain stable. The long-lived stable phase in low-$q$ systems is enabled by orbital widening once the mass ratio is less than 1.0, which maintains Roche lobe contact without driving the system toward instability.

\item In contrast, for systems with higher mass ratios, mass transfer grows exponentially and may lead to the onset of a common-envelope phase. The determination of the critical mass ratio is the subject of an independent study and is not part of this work. For mass ratios in the range $q=2-4$ (depending on period and donor mass), the rapid orbital shrinkage drives a steep rise in the mass transfer rate, reducing the symbiotic-phase duration to only $\sim10^4$–$10^5$ yr before the system approaches $L_2$ overflow, still well above previously assumed values of $\sim10^3$ yr.

\item Overall, for both RGB and AGB donors, including stellar winds and other non-conservative mass and angular-momentum loss processes is expected to slow orbital shrinkage and delay the onset of rapid mass transfer. The time spent in the adopted SySt mass-transfer-rate range should therefore be regarded as closer to a lower limit. For AGB donors, the SySt phase is episodic and occurs only during thermal pulses. The caveat is that stellar winds can remove the AGB envelope and reduce the total number of thermal pulses. However, in our models the AGB donors usually lose their envelopes through Roche-lobe overflow within the first few thermal-pulse episodes, so this effect is unlikely to substantially decrease the SySt durations reported here. The final orbital periods and mass ratios should be interpreted within the fully conservative assumption.
\end{itemize}

\item Our models successfully reproduce the observed properties of five well-characterized symbiotic systems. In particular, the evolutionary tracks match both typical systems and recurrent novae, reproducing their observed donor and accretor masses, orbital periods, and mass ratios within the observational uncertainties. This close agreement demonstrates that our binary evolution models can reliably describe the observed population of symbiotic binaries.

\end{enumerate}

\par \paragraph{Comparison with observed properties:}  
The models naturally produce donors with Roche-lobe filling factors consistent with the ellipsoidal variability observed in many systems, and the predicted orbital period evolution places long-lived stable binaries in the observed $P\sim 100$–1100\,day range. 
The evolutionary tracks cover the full observed region in the $P$–$q$ plane and reproduce both ordinary S-type symbiotic systems and recurrent novae. 
The adopted donor-mass range of 1–3\,$M_\odot$ matches that inferred for S-type systems. 
Long-lived systems originate primarily from RGB donors, whereas AGB donors contribute only short episodic phases during thermal pulses. 
Our results apply to S-type systems; D-type/Mira binaries and WRLOF channels lie outside the scope of this work.

In conclusion, our study provides a framework for understanding the mass transfer mechanisms in SySts and helps bridge the gap between theoretical predictions and observational evidence. Because our models assume fully conservative mass transfer, they provide the shortest possible symbiotic lifetimes; any realistic non-conservative mass would only extend these timescales further. The use of RUMT not only explains the stability of S-type symbiotic systems but also offers new insights into their possible evolutionary pathways and outcomes. In future work, we plan to employ binary population synthesis to further investigate the distribution of SySts.

\begin{acknowledgements}
 We thank the anonymous referee for his/her constructive comments and suggestions that helped improve the manuscript. We are grateful to Jingxiao Luo, Lifu Zhang, and Boyang Guo for their fruitful discussions. 
 This work is supported by the National Natural Science Foundation of China (NSFC Nos. 12125303, 12288102, 12090040/3, 12090040/1, 12333008), the National Key R\&D Program of China (grant Nos. 2021YFA1600403, 2021YFA1600401 and 2021YFA1600400), Yunnan Fundamental Research Projects (grant Nos. 202401BC070007, 202201BC070003 and 202001AW070007), International Centre of Supernovae (ICESUN), Yunnan Key Laboratory of supernova Research (No. 202505AV340004) and the Yunnan Revitalization Talent Support Program — Science \& Technology Champion Project (No. 202305AB350003). This work has also been supported by the New Cornerstone Science Foundation through the XPLORER PRIZE, the Strategic Priority Research Program of the Chinese Academy of Sciences, the Yunnan Revitalization Talent Support Program "YunLing Scholar", and the CAS Project for Young Scientists in Basic Research (YSBR-148). N.I. acknowledges funding from NSERC under Discovery grant No. RGPIN-2025-05603. JM was supported by the Polish National Science Centre (NCN) grant 2023/48/Q/ST9/00138. The authors gratefully acknowledge the ``PHOENIX Supercomputing Platform'' jointly operated by the Group of Binary Population Synthesis and the Group of Stellar Astrophysics at Yunnan Observatories, Chinese Academy of Sciences.

\end{acknowledgements}

\bibliographystyle{aa}
\bibliography{aa}

\begin{appendix}
\section{Orbital and stellar parameters of selected symbiotic binaries}

\noindent In Table \ref{tab:observation}, we present the information for the selected observational objects. In this table, we list their orbital periods $P_\mathrm{orb}$, mass function $f(m_\mathrm{g})$, mass ratio $q$, estimated giant mass $M_\mathrm{g}$, and WD accretor mass $M_\mathrm{WD}$. The column “ELC” indicates whether the system exhibits an ellipsoidal light curve.

\section{HR diagrams: comparison of observations and models}

\noindent In addition to the orbital parameters, we include the luminosity of the cool giant $L_\mathrm{g}$ and its effective temperature $\log T_\mathrm{eff}$, as well as chemical information whenever available in Table~\ref{tab:observation2} \citep{galan2015chemical,galan2016chemical,galan2017chemical,Ga_an_2023}. These chemical data comprise metallicity ([Fe/H]), the chemical abundance $^{12}\mathrm{C}/^{13}\mathrm{C}$, and the presence of $s$-process element enrichment. The low $^{12}\mathrm{C}/^{13}\mathrm{C}$ ratios and the enrichment in $^{14}$N observed in all studied objects indicate that the symbiotic giants have experienced the first dredge-up. The presence of $s$-process elements may have either an intrinsic or extrinsic origin. To further constrain the evolutionary status of the donor stars, we also note the presence of low-amplitude pulsations: periods of $\gtrsim 70$ days are characteristic of early-AGB stars, whereas periods of $\sim 50$–70 days may correspond to either RGB or early-AGB stars \citep{2013AcA....63..405G}. The last column of the table lists the luminosity of the hot component $L_\mathrm{WD}$ \citep{mikolajewska2011symbioticnovae}. These values are typically uncertain within a factor of $\sim 2$, but they suggest accretion rates above $\sim 10^{-8}\,M_\odot\,\mathrm{yr}^{-1}$. In many systems, the hot component luminosity is consistent with stable thermonuclear burning of the accreted hydrogen.

\par In Figure~\ref{HR}, we show the Hertzsprung–Russell diagram of our models to compare with the observational objects. Most of the giants occupy the region near the RGB tip or early-AGB phase. For systems with $M_\mathrm{d,i}=2.0$–$3.0,M_\odot$, the evolutionary tracks on the HR diagram deviate from the observational data, with the observed systems exhibiting systematically lower effective temperatures and luminosities than the models.

\section{Evolutionary track for simulated models}
\begin{figure*}[htbp]
    \centering
    \begin{minipage}{0.49\linewidth}
        \centering
        \includegraphics[width=0.9\linewidth]{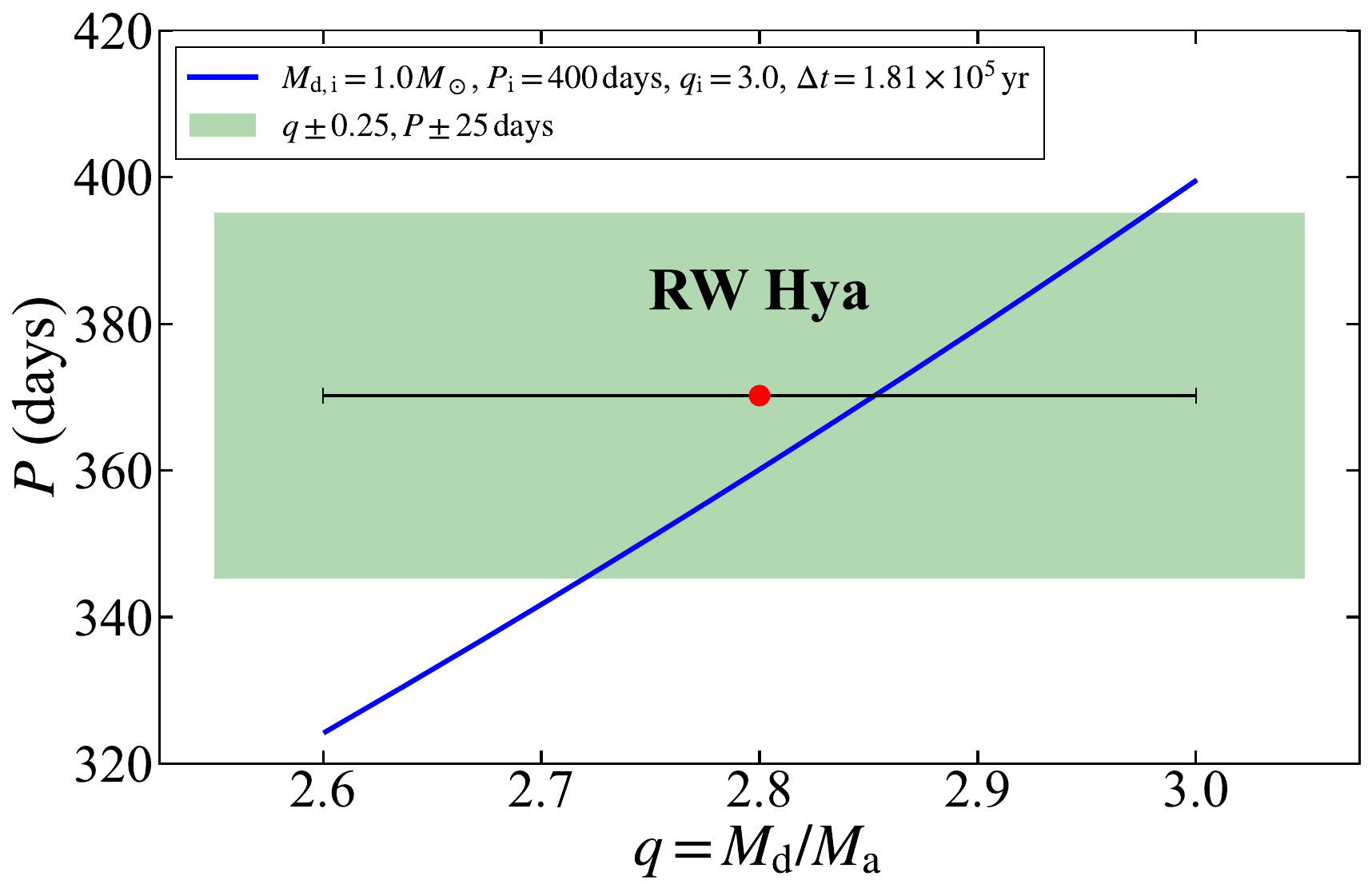}
        \label{exact_1}
    \end{minipage}
    \begin{minipage}{0.49\linewidth}
        \centering
        \includegraphics[width=0.9\linewidth]{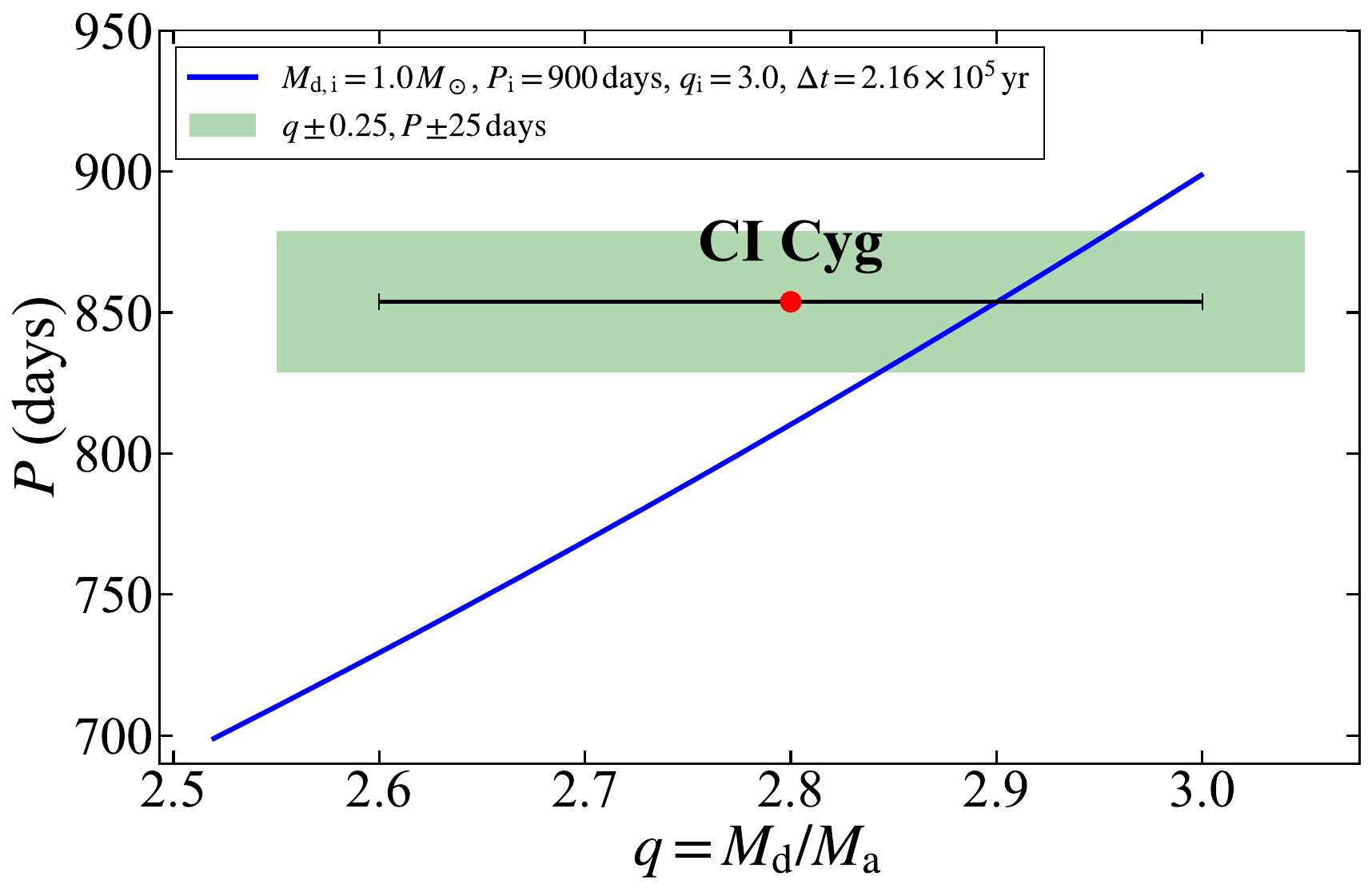}
        \label{exact_2}
    \end{minipage}

    \vspace{0.5cm}

    \begin{minipage}{0.49\linewidth}
        \centering
        \includegraphics[width=0.9\linewidth]{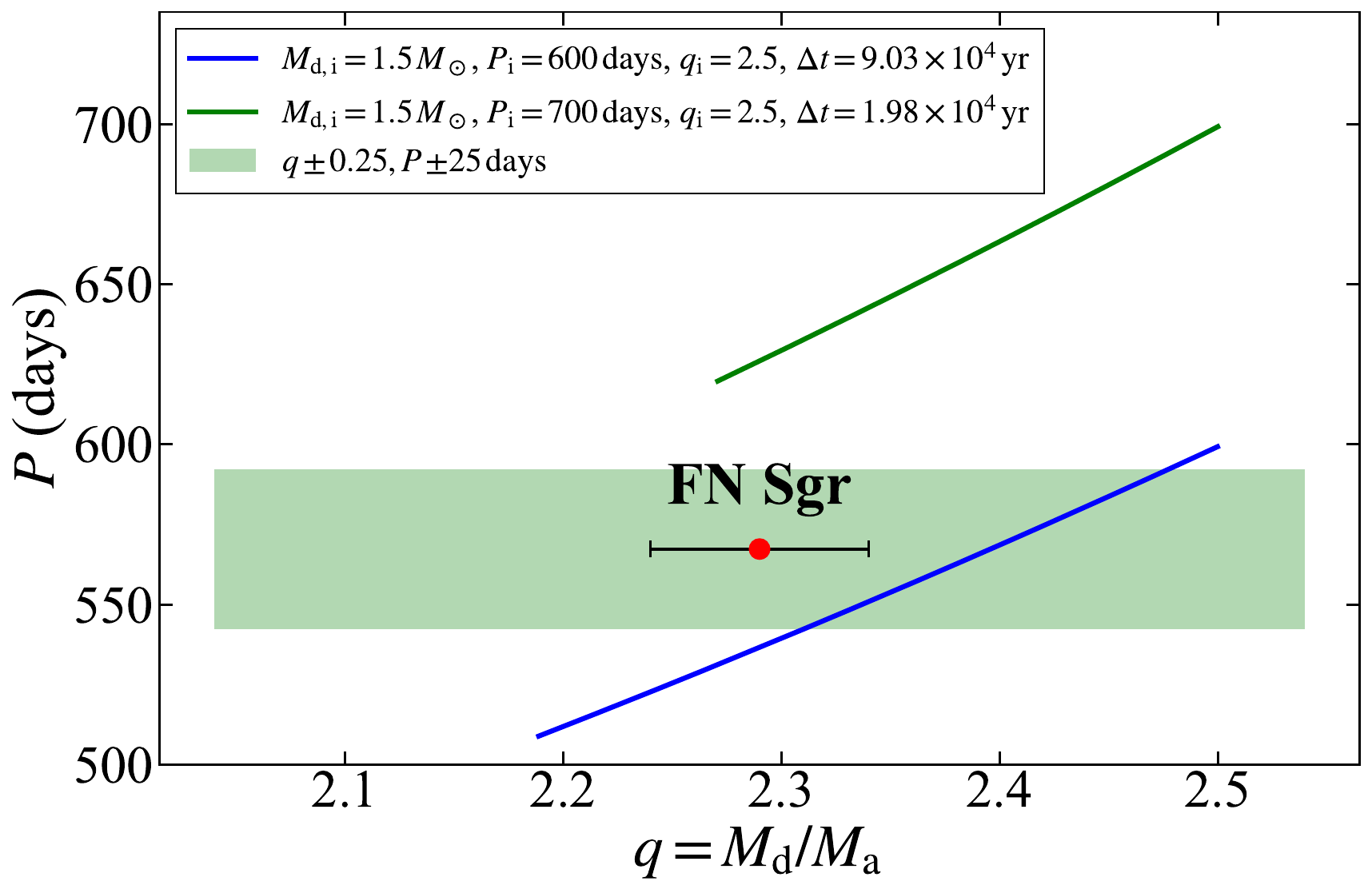}
        \label{exact_3}
    \end{minipage}
    \begin{minipage}{0.49\linewidth}
        \centering
        \includegraphics[width=0.9\linewidth]{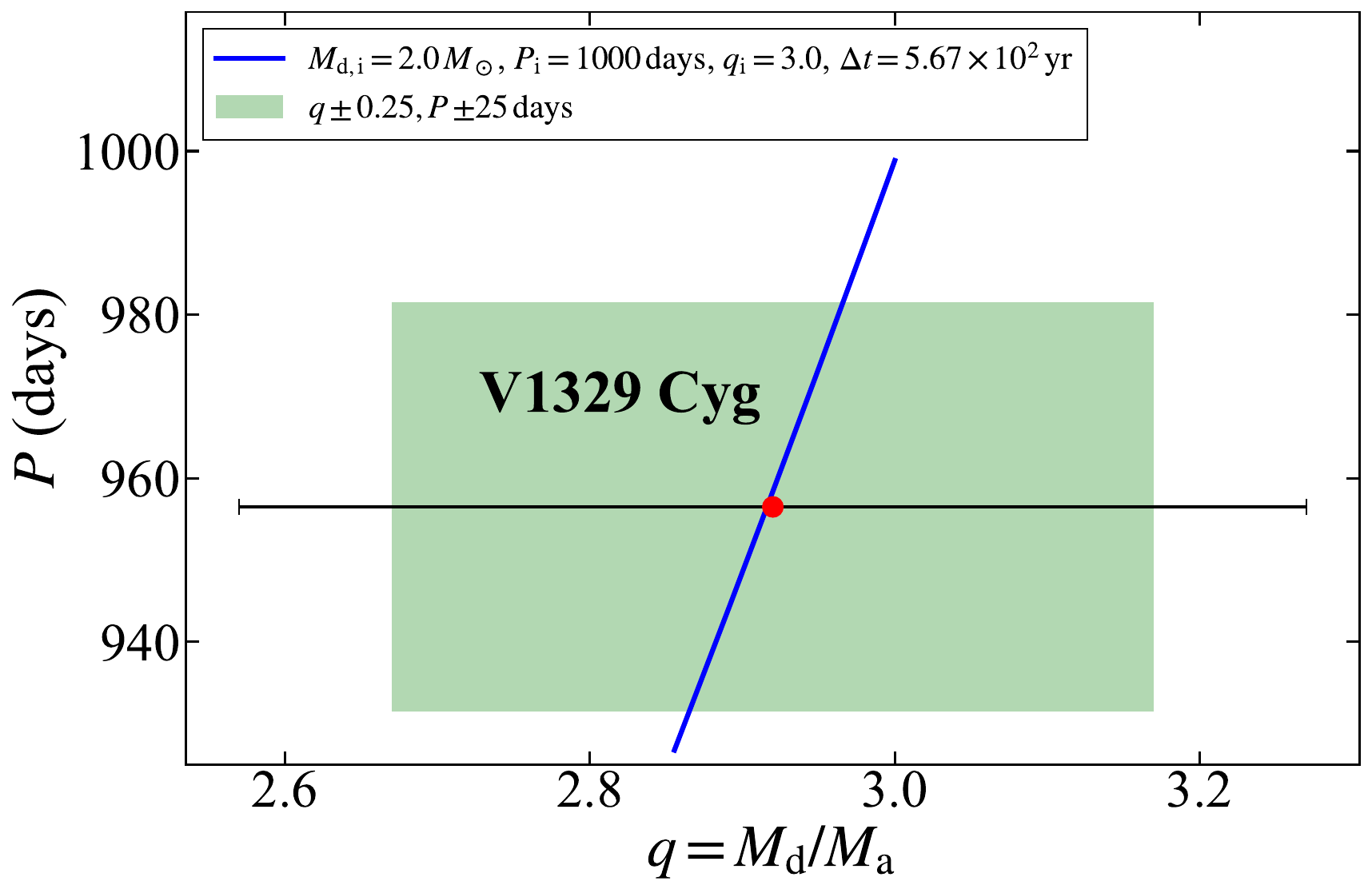}
        \label{exact_4}
    \end{minipage}

    \vspace{0.5cm}

    \begin{minipage}{0.49\linewidth}
        \centering
        \includegraphics[width=0.9\linewidth]{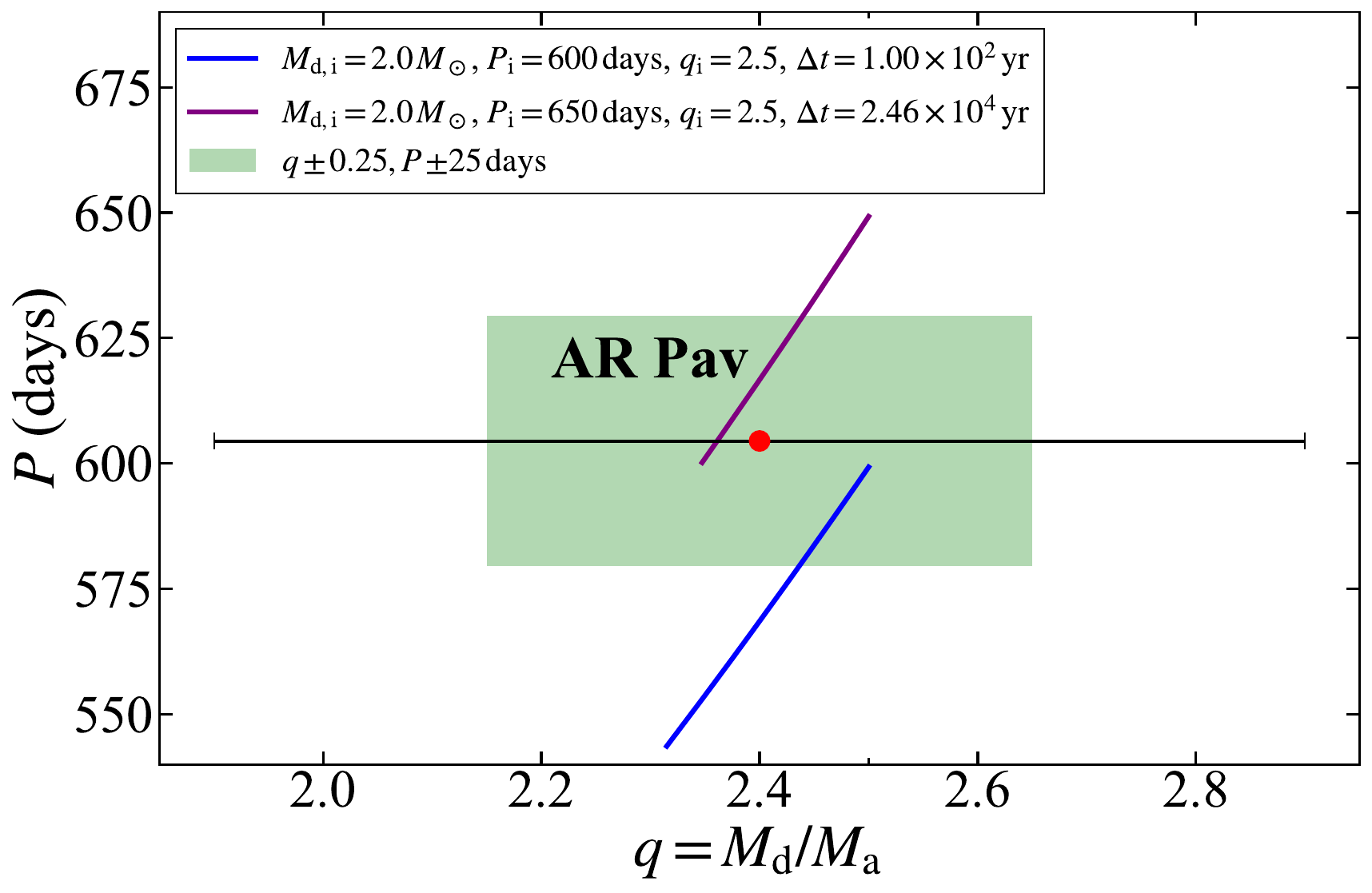}
        \label{exact_5}
    \end{minipage}

    \caption{Comparison between our model predictions and the observational systems which are observationally well-determined. The green shaded area represents the criteria which
the orbital period within 50 days, donor mass within 0.25 $M_\odot$. The colored trajectories represent the evolutionary tracks from our models.}
    \label{8}
\end{figure*}
\noindent In Figure~\ref{evolutionary_sequence}, we show the mass transfer rate as a function of time for all models. The initial donor mass $M_\mathrm{d,i}$ and orbital period $P_\mathrm{i}$ are labeled in the figure, while different colors and symbols denote different initial mass ratios $q_\mathrm{i}$. For models with $M_\mathrm{d,i}=1.0,M_\odot$ and $P_\mathrm{i}=300$–700 days, the tracks exhibit small fluctuations, which may arise from the layer calculation procedure in MESA.

\clearpage

\onecolumn
\begin{center}
\begin{minipage}{\textwidth}
\captionof{table}{Orbital and stellar parameters for selected symbiotic stars. }
\label{tab:observation}
\end{minipage}
\end{center}

\vspace{-4em}  
\begin{longtable}{@{}llrrlllp{4.0cm}l@{}}

 \\
\toprule
No. & Name & $P_{\mathrm{orb}}$ & $f(m_{\rm g})$ & $q$ & $M_{\mathrm{g}}$ & $M_{\mathrm{WD}}$ & Reference & ELC \\
 &  & (days) &  &  & ($M_\odot$) & ($M_\odot$) &  &  \\
\midrule
\endfirsthead

\toprule
No. & Name & $P_{\mathrm{orb}}$ & $f(m_{\rm g})$ & $q$ & $M_{\mathrm{g}}$ & $M_{\mathrm{WD}}$ & Reference & ELC \\
 &  & (days) &  &  & ($M_\odot$) & ($M_\odot$) &  &  \\
\midrule
\endhead

\midrule
\multicolumn{9}{r}{\textit{Continued on next page}} \\
\midrule
\endfoot

\bottomrule
\endlastfoot

3 & EG And & 483.3 & 0.0200 &   & 1.5 & 0.4–0.7 & \cite{kenyon2016eg} & yes \\
8 & AX Per & 682.1 & 0.0333 & 2.4±0.4 & ~1.1 & ~0.46 & \cite{fekel2000infrared};\newline \cite{1992AJ....103..579M} & yes \\
11 & BD Cam & 596.2 & 0.0370 &   & ~1 & ~0.4–0.7 & \cite{griffin1984spectroscopic}& no \\
17 & V1261 Ori & 638.24 & 0.0320 &   & ~1 & ~0.5–0.8 & \cite{boffin2014roche};\newline\cite{2025AA...695A..61M}& yes \\
23 & BX Mon & 1276.0 & 0.0103 & 4.4±0.5 & ~1.5 & ~0.34 & Mikolajewska et al. - in prep. & yes \\
31 & Hen 3-461 & 2271.0 & 0.0860 &   & ~1.5 & ~0.76–1.37 & \cite{2017AJ....153...35F} & no \\
33 & SY Mus & 625.0 & 0.0429 & ~2.6 & ~1.3 & ~0.5 & \cite{2017AJ....153...35F};\newline\cite{2007BaltA..16...49R} & yes \\
36 & TX CVn & 199.0 & 0.0040 &   & ~1 & >~0.3 & \cite{1989AJ.....97..194K} & no \\
38 & Hen 3-828 & 660.5 & 0.0340 &   & ~1.5 & ~0.5–0.9 & \cite{2017AJ....153...35F} & yes \\
41 & Sa 22/St 2-22 & 918.0 & 0.0331 & 3.5±0.5 & 2.8±0.7 & 0.8±0.2 & \cite{galan2022symbiotic} & no \\
42 & CD-36 8436 & 985.0 & 0.0083 &   & ~1.5 & ~0.3–0.5 & \cite{2015AJ....150...48F} & no \\
45 & RW Hya & 370.2 & 0.0260 & ~3 & 1.0±0.2 & 0.36±0.10 & \cite{Kenyon1995};\newline\cite{Schild1996};\newline\cite{2007BaltA..16...49R};\newline\cite{2025AA...695A..61M}& yes \\
51 & BD-21 3873 & 281.6 & 0.0350 &   & ~0.9 & ~0.5 & \cite{smith1997bd}& yes \\
57 & T CrB & 227.549 & 0.3295 & 0.50 & 0.69±0.02 & 1.37±0.01 & \cite{hinkle2025binary} & yes \\
58 & AG Dra & 549.0 & 0.0115 &   & ~1–1.2 & ~0.44–0.49 & \cite{fekel2000infrared} & no \\
65 & Hen 3-1213 & 533.0 & 0.0157 &   & ~1 & ~0.3–0.5 & \cite{2015AJ....150...48F} & yes \\
66 & Hen 2-173 & 911.0 & 0.0517 &   & ~1.5 & ~0.6–1.1 & \cite{fekel2006infrared} & yes \\
68 & KX Tra & 1350.0 & 0.0390 & ~2.9 & ~1.7–2.6 & ~0.6–0.9 & \cite{ferrer2003symbiotic} & no \\
71 & CL Sco & 625.0 & 0.0209 &   & ~1.5 & ~0.4–0.7 & \cite{fekel2006infrared} & no \\
73 & V455 Sco & 1398.0 & 0.0724 &   & ~1.1 & ~0.6 & \cite{fekel2008infrared} & no \\
88 & M1-21 & 898.0 & 0.0652 &   & ~1.2 & ~0.6 & \cite{fekel2008infrared}& no \\
93 & AE Ara & 803.0 & 0.0235 &   & ~1.6 & ~0.46–0.79 & \cite{fekel2010infrared} & yes \\
94 & SS73 96 & 828.2 & 0.0547 &   & ~1.5 & >~0.63 & \cite{2015AJ....150...48F}& no \\
101 & RS Oph & 453.6 & 0.2540 & 0.59±0.05 & 0.68–0.8 & 1.2–1.4 &\cite{brandi2009spectroscopic}& no \\
119 & AS 270 & 671.2 & 0.0189 &   & ~1.5 & ~0.4–0.7 & \cite{fekel2006infrared}& yes \\
131 & Y CrA & 1619.0 & 0.0275 &   & ~1–1.5 & ~0.4–0.8 & \cite{fekel2010infrared}& no \\
134 & FG Ser & 633.5 & 0.0220 &   & ~2 & ~0.52–0.87 & \cite{fekel2000infrared}& yes \\
136 & Hen 2-374 & 820.0 & 0.0416 &   & ~1.6 & ~0.6–1.0 & \cite{fekel2010infrared} & no \\
142 & AR Pav & 604.5 & 0.0696 & 2.4 & 1.9–2.3 & 0.80–0.97 & \cite{quiroga2002spectroscopic};\newline\cite{2017AJ....153...35F}& yes \\
145 & V443 Her & 599.4 & 0.0010 &   & ~2 & >0.3 & \cite{fekel2000infrared};\newline\cite{1993AJ....106..284D}& no \\
148 & V3890 Sgr & 747.6 & 0.3360 & 0.78±0.05 & 1.05±0.11 & 1.35±0.13 & \cite{mikolajewska2021symbiotic} & no \\
156 & FN Sgr & 568.3 & 0.0589 & 2.2 & 1.33±0.24 & 0.60±0.09 & \cite{brandi2005spectroscopic};\newline\cite{belloni2024formation}& yes \\
166 & BF Cyg & 757.2 & 0.0239 & ~3.6 & ~2 & ~0.6 & \cite{fekel2001infrared};\newline\cite{yudin2005ir}& yes \\
167 & CH Cyg & 5689.0 & 0.0510 &   & ~2 & ~0.8 & \cite{hinkle2009infrared} & no \\
170 & Hen 3-1761 & 562.0 & 0.0230 & 4.2 & >~2.5 & >~0.6 & \cite{2006BAAA...49..132B} & no \\
172 & CI Cyg & 853.8 & 0.0260 & 2.8±0.2 & 1.20±0.22 & 0.43±0.08 & \cite{1991AJ....101..637K};\newline\cite{fekel2000infrared};\newline\cite{mikolajewska2006line}& yes \\
180 & ER Del & 2094.2 & 0.0770 &   & 2±0.4 & >~0.85±0.10 & \cite{boffin2014roche};\newline\cite{2025AA...695A..61M};\newline\cite{jorissen2019barium}& no \\
181 & V1329 Cyg & 956.5 & 0.0481 & 2.9 & 2.1±0.7 & 0.74±0.08 & \cite{fekel2001infrared};\newline\cite{2003ASPC..303....9M}& no \\
182 & CD-43 14304 & 1442.0 & 0.0139 &   &   &   & \cite{schmid1998high}& no \\
185 & AG Peg & 816.5 & 0.0135 & 2.6–5.8 & ~2 & ~0.5 & \cite{1993ApJ...407L..81K};\newline\cite{fekel2000infrared};\newline\cite{2025AA...695A..61M}& no \\
187 & Z And & 758.8 & 0.0240 &   & ~1.5 & ~0.76 & \cite{fekel2000infrared}& yes \\
188 & R Aqr & 15470.0 & 0.0960 & ~1.4 & ~1 & ~0.7 & \cite{gromadzki2009spectroscopic};\newline \cite{2023hsa..conf..190A} & no \\
s11 & CD-27 8661 & 751.8 & 0.1100 &   & ~1–1.5 & >~0.68–0.85 & \cite{2012BaltA..21...39J};\newline\cite{jorissen2019barium} & yes \\
*new & DASCH J0757 & 119.2 & 0.0331 &   & 1.1±0.3 & 0.6±0.2 & \cite{tang2012dasch} & yes \\
\end{longtable}
\noindent\textit{Notes.} Orbital and stellar parameters for selected symbiotic stars. The first column gives the identification numbers adopted from \citet{2000A&AS..146..407B}. All systems in this sample have well-determined orbital periods $P_\mathrm{orb}$, mass function $f(m_\mathrm{g})$, mass ratio $q$, estimated giant mass $M_\mathrm{g}$ and WD accretor mass $M_\mathrm{WD}$.
\twocolumn

\onecolumn
\begin{center}
\begin{minipage}{\textwidth}
\captionof{table}{Stellar parameters of the selected observational symbiotic systems.}
\end{minipage}
\end{center}
\centering
\begin{longtable}{@{}llrrllll@{}}
\label{tab:observation2} \\
\toprule
No. & Name & $L_\mathrm{g}$ & $\log T_\mathrm{eff}$ & $L_\mathrm{WD}$ & [Fe/H] & $^{12}\mathrm{C}/^{13}\mathrm{C}$ & s-process \\ 
 & & ($L_\odot$) & & ($L_\odot$) & & & \\
\midrule
\endfirsthead
\toprule
No. & Name & $L_\mathrm{g}$ & $\log T_\mathrm{eff}$ & $L_\mathrm{WD}$ & [Fe/H] & $^{12}\mathrm{C}/^{13}\mathrm{C}$ & s-process \\ 
 & & ($L_\odot$) & & ($L_\odot$) & & & \\
\midrule
\endhead
\midrule
\multicolumn{8}{r}{\textit{Continued on next page}} \\
\midrule
\endfoot
\bottomrule
\endlastfoot
3 & EG And & 3.26 & 3.546 & 26-52 & -0.54 & 7 & no \\
8 & AX Per & 3.2 & 3.5185 & 400-2300 & -0.26 & 9.5 & no \\
11 & BD Cam & 3.36 & 3.556 &  &  &  & yes \\
17 & V1261 Ori & 3.05 & 3.556 &  &  &  & yes \\
23 & BX Mon & 3.26 & 3.5315 & 195 & -0.40 & 8 & no \\
31 & Hen 3-461 & 3.45 & 3.5051 &  & 0.12 & 13 & no \\
33 & SY Mus & 3.38 & 3.5315 & 3160 & -0.15 & 8 & no \\
36 & TX CVn & 3.12 & 3.613 & 240 &  &  & ? \\
38 & Hen 3-828 & 3.36 & 3.5185 & 1560 & 0.03 & 15 & no \\
41 & Sa 22/St 2-22 & 3.21 & 3.5315 & 1180 &  &  & no \\
42 & CD-36 8436 & 3.25 & 3.5185 &  & -0.30 & 8 & no \\
45 & RW Hya & 3.23 & 3.5682 & 1400 & -0.77 & 5.3 & no \\
51 & BD-21 3873 & 3.10 & 3.5776 & 2000 & -1.3 &  & yes \\
57 & T CrB & 2.75 & 3.5315 &  & 0.35 &  & no \\
58 & AG Dra & 3.23 & 3.5894 & 1170 & -1.3 &  & yes \\
65 & Hen 3-1213 & 3.30 & 3.6128 & 1170 & -0.68 &  & ? \\
66 & Hen 2-173 & 3.52 & 3.5315 & 2270 & -0.18 &  & no \\
68 & KX Tra & 3.55 & 3.5185 & 15500 & -0.34 &  & no \\
71 & CL Sco & 3.70 & 3.5315 & 2170 & -0.31 &  & no \\
73 & V455 Sco & 3.25 & 3.5051 & 1860 & 0.36 &  & no \\
88 & M1-21 & 3.56 & 3.5185 & 3450 &  &  & no \\
93 & AE Ara & 3.25 & 3.5185 & 2000 & -0.02 &  & no \\
94 & SS73 96 & 3.17 & 3.5315 &  & -0.24 &  & no \\
101 & RS Oph & 3.06 & 3.626 &  &  &  & no \\
119 & AS 270 & 3.46 & 3.5185 & 400 & 0.03 &  & no \\
131 & Y CrA & 3.17 & 3.5185 & 190 & -0.40 &  & no \\
134 & FG Ser & 3.36 & 3.5315 & 720-1130 & -0.08 &  & no \\
136 & Hen 2-374 & 3.17 & 3.5185 & 1750 & -0.52 &  & no \\
142 & AR Pav & 3.34 & 3.5315 & 4300-10000 & -0.26 &  & no \\
145 & V443 Her & 3.44 & 3.5185 & 1650 & -0.02 &  & no \\
148 & V3890 Sgr & 3.41 & 3.5051 &  &  &  & no \\
156 & FN Sgr & 3.54 & 3.5502 & 1500-3200 &  &  & no \\
166 & BF Cyg & 3.55 & 3.5315 & 9500 & -0.25 & 6.1 & no \\
167 & CH Cyg & 3.56 & 3.5051 &  & 0.13 &  & no \\
170 & Hen 3-1761 & 3.50 & 3.5185 & 380 & -0.25 &  & yes \\
172 & CI Cyg & 3.56 & 3.5185 & 1000-4300 & -0.10 & 12.6 & no \\
180 & ER Del & 3.27 & 3.5315 &  &  &  & yes \\
181 & V1329 Cyg & 3.41 & 3.5051 & 19700 & 0.12 & 24 & no \\
182 & CD-43 14304 & 3.50 & 3.5911 &  & -0.93 &  & yes \\
185 & AG Peg & 3.41 & 3.5502 & 4800-1500 & -0.51 & 5.2 & no \\
187 & Z And & 3.35 & 3.5315 & 900-4300 & -0.06 & 10.5 & no \\
188 & R Aqr &  & Mira variable &  &  &  & no \\
s11 & CD-27 8661 & 3.48 & 3.5315 &  &  &  & yes \\
*new & DASCH J0757 & 2.59 & 3.5627 &  &  &  & no \\
\end{longtable}
\noindent\textit{Notes.} The table lists the luminosity $L_\mathrm{g}$ and effective temperature $\log T_\mathrm{eff}$ of the cool giant, its chemical properties ([Fe/H], $^{12}\mathrm{C}/^{13}\mathrm{C}$ ratio, and $s$-process enrichment), and the luminosity of the hot component $L_\mathrm{WD}$.
\twocolumn

\begin{figure*}[htbp]
    \centering
    \begin{minipage}{0.95\linewidth}
        \centering
        \includegraphics[width=0.95\linewidth]{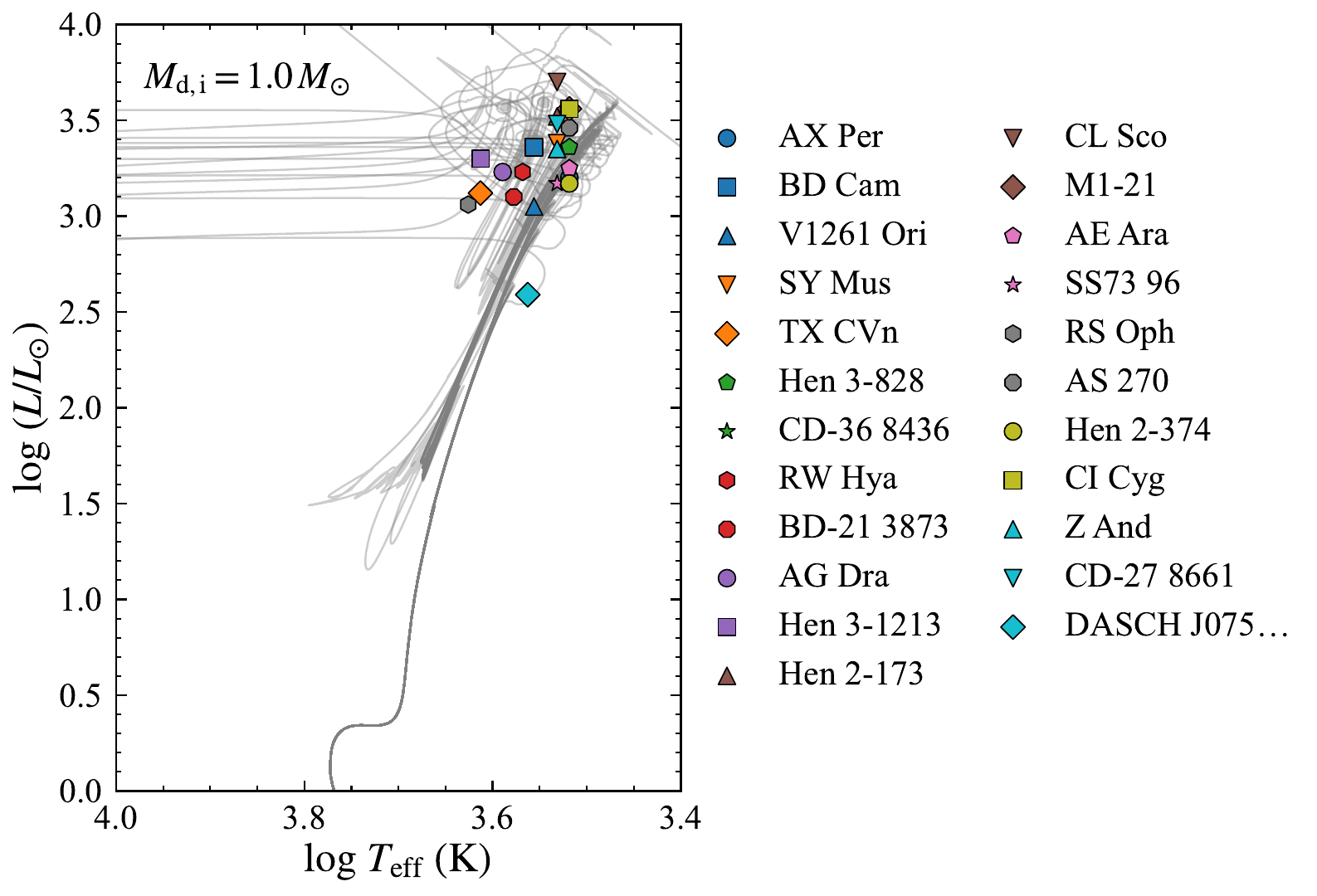}
        \label{fig:compare_1.0}
    \end{minipage}
    \begin{minipage}{0.95\linewidth}
        \centering
        \includegraphics[width=0.95\linewidth]{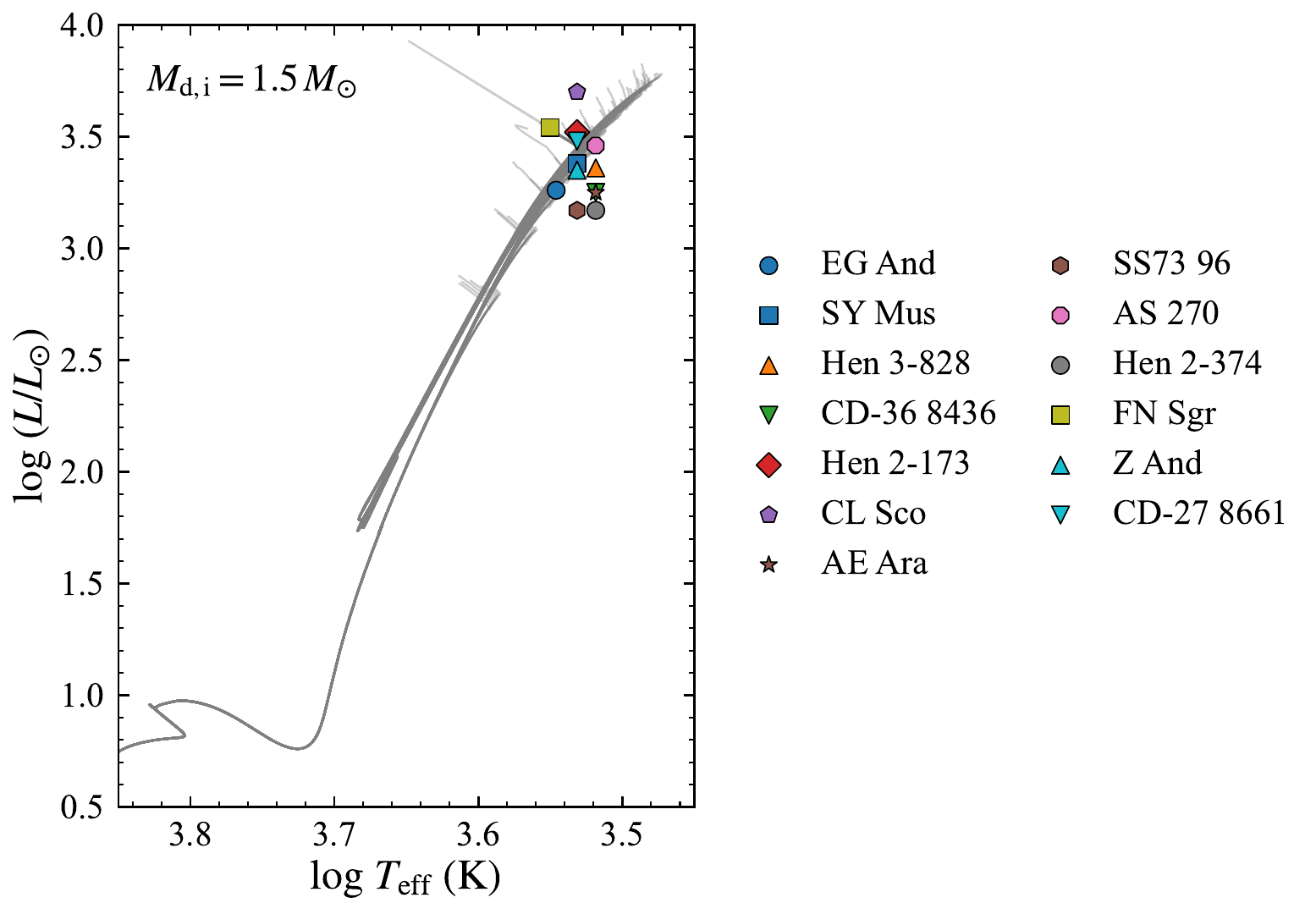}
        \label{fig:compare_1.5}
    \end{minipage}

    \caption{HR diagrams of the observational systems and our models. Different symbols with different colors represents different observational samples. The gray line represents the evolutionary track of our models. (Continue on next page)}
    \label{HR}
\end{figure*}

\begin{figure*}[htbp]
    \ContinuedFloat 
    \centering

    \begin{minipage}{0.95\linewidth}
        \centering
        \includegraphics[width=0.95\linewidth]{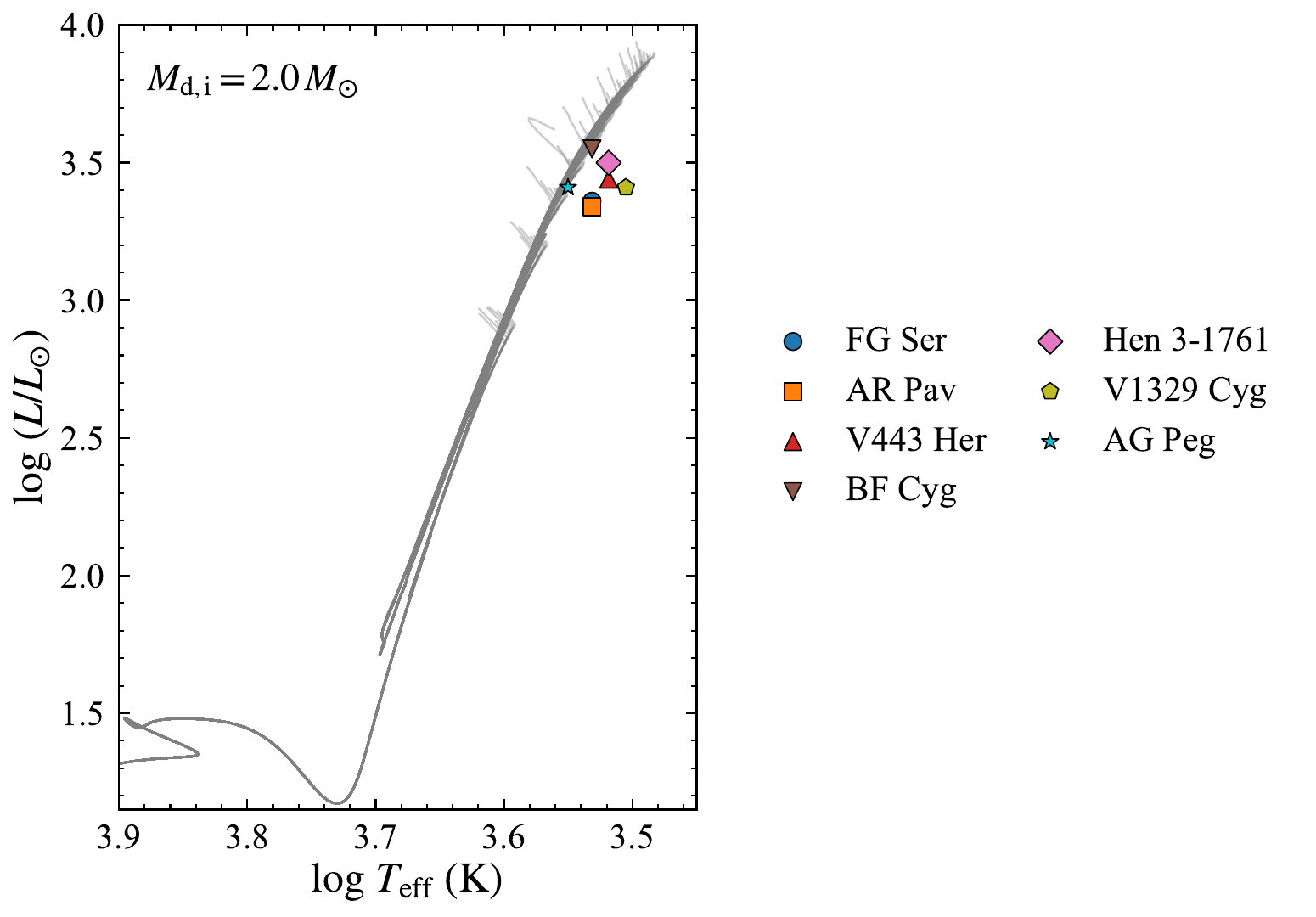}
        \label{fig:compare_2.0}
    \end{minipage}
    \begin{minipage}{0.95\linewidth}
        \centering
        \includegraphics[width=0.95\linewidth]{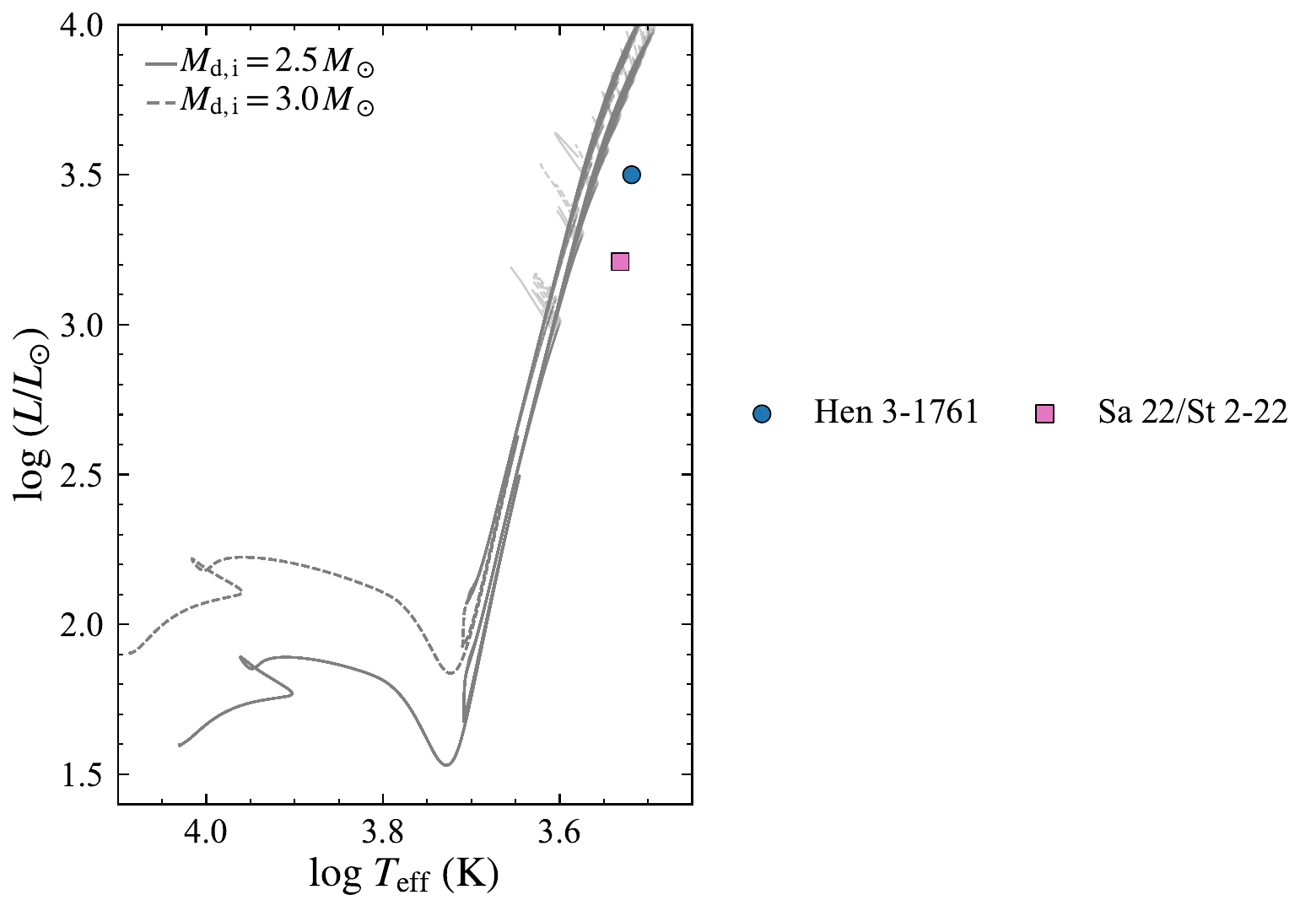}
        \label{fig:compare_2.5}
    \end{minipage}

    \caption{HR diagrams of the observational systems and our models. Different symbols with different colors represents different observational samples. The gray line represents the evolutionary track of our models. (Continued)} 
\end{figure*}

\twocolumn

\begin{figure*}[p]
    \centering
    \includegraphics[width=0.9\textwidth]{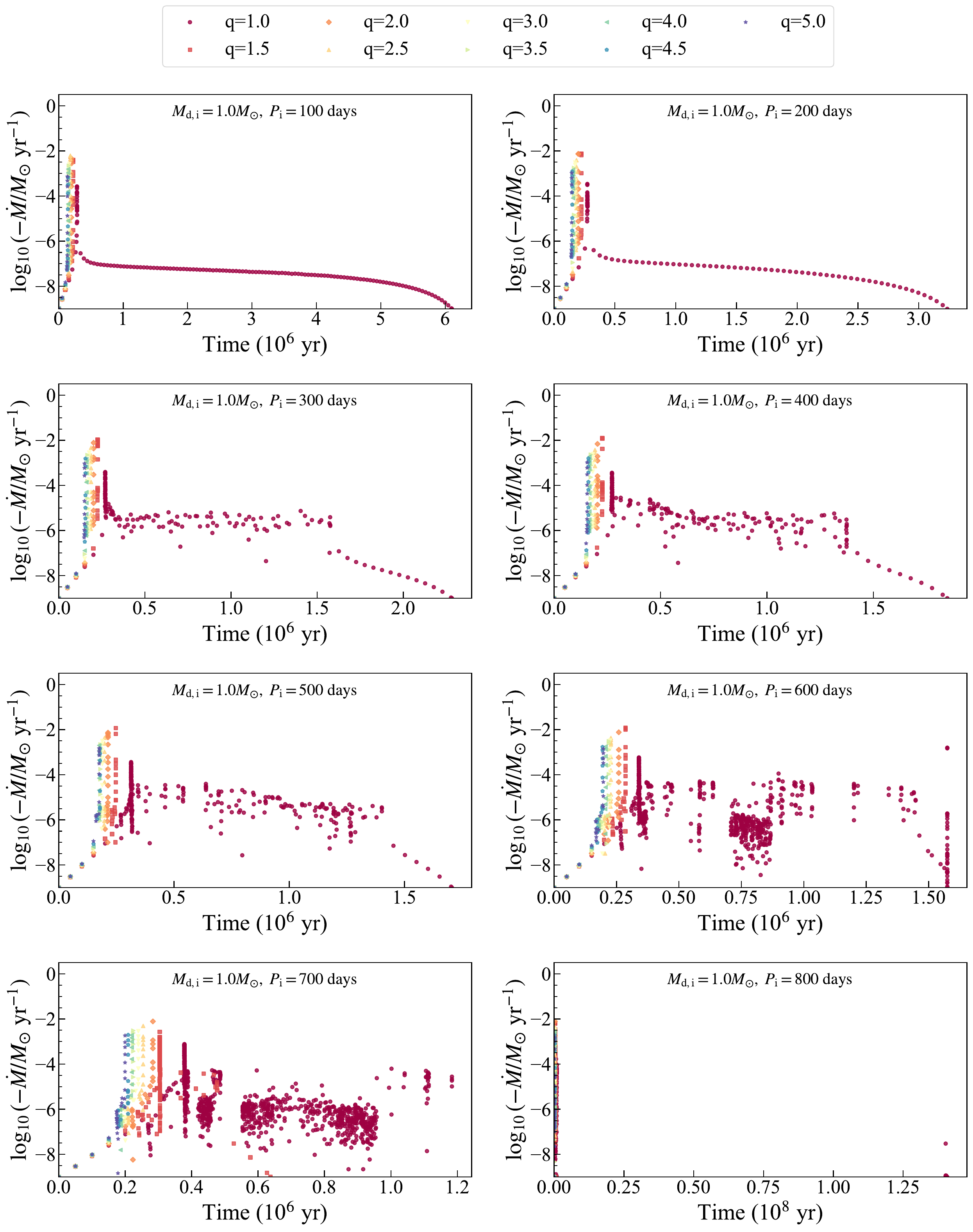}
    \caption{Evolution of the mass transfer rate for all models in our grid. Each panel corresponds to a specific initial donor mass ($M$) and orbital period ($P$). (Continued on next page)}
    \label{evolutionary_sequence}
\end{figure*}
\clearpage

\begin{figure*}[p]
    \ContinuedFloat
    \centering
    \includegraphics[width=0.9\textwidth]{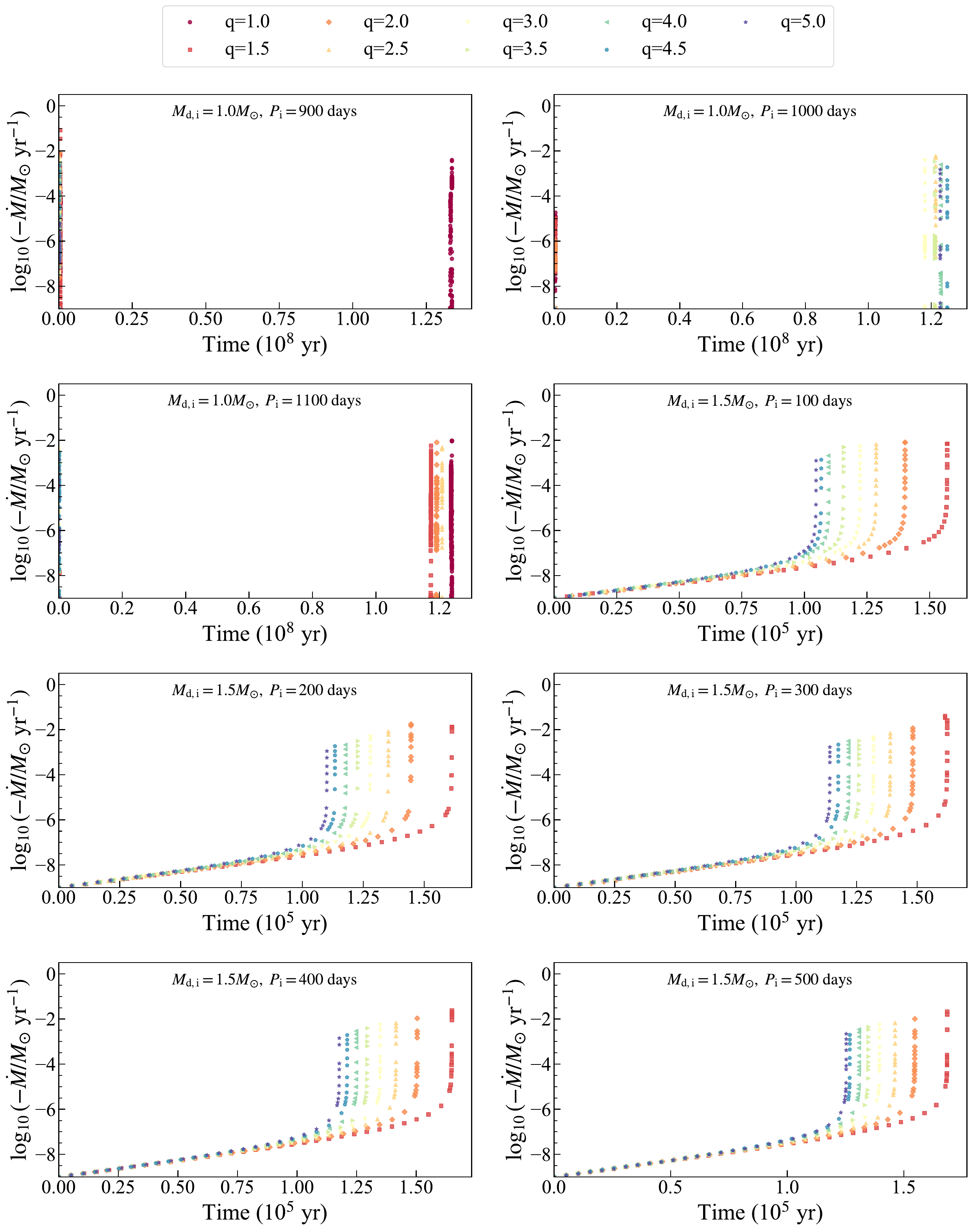}
    \caption{Evolution of the mass transfer rate for all models in our grid. Each panel corresponds to a specific initial donor mass ($M_\mathrm{d,i}$) and orbital period ($P_\mathrm{i}$). (Continued)}
\end{figure*}
\clearpage

\begin{figure*}[p]
    \ContinuedFloat
    \centering
    \includegraphics[width=0.9\textwidth]{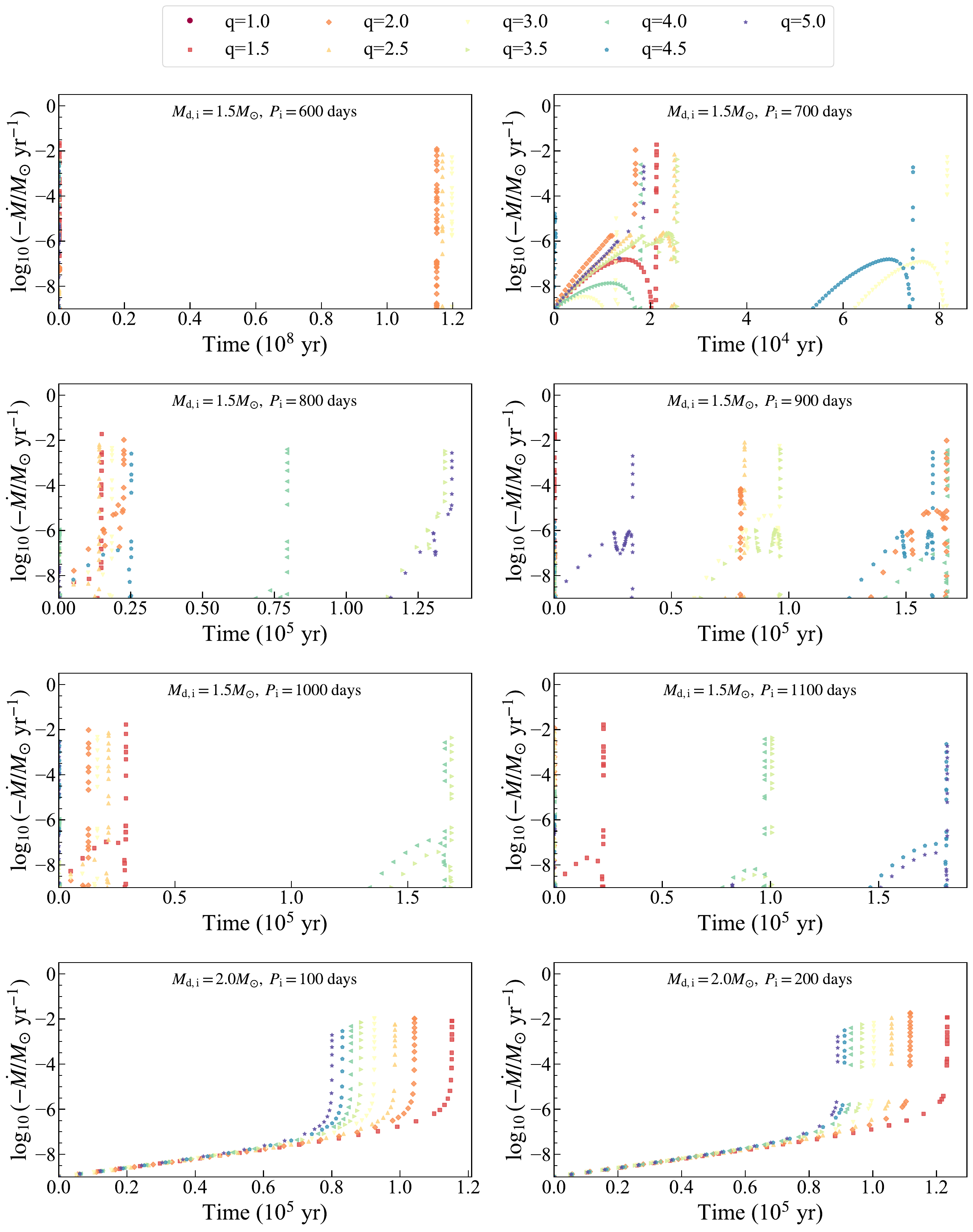}
    \caption{Evolution of the mass transfer rate for all models in our grid. Each panel corresponds to a specific initial donor mass ($M$) and orbital period ($P$). (Continued)}

\end{figure*}
\clearpage

\begin{figure*}[p]
    \ContinuedFloat
    \centering
    \includegraphics[width=0.9\textwidth]{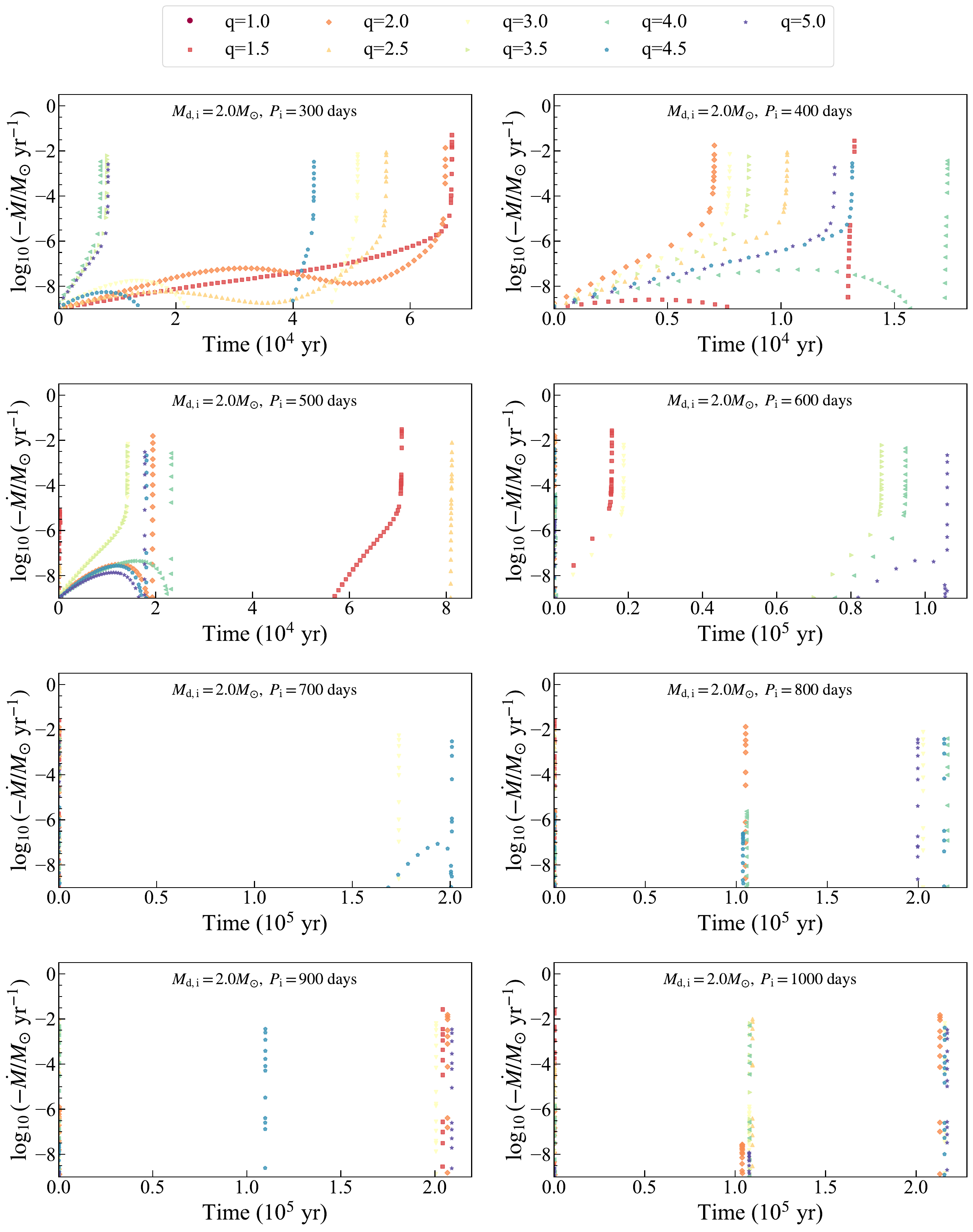}
    \caption{Evolution of the mass transfer rate for all models in our grid. Each panel corresponds to a specific initial donor mass ($M$) and orbital period ($P$). (Continued)}

\end{figure*}
\clearpage

\begin{figure*}[p]
    \ContinuedFloat
    \centering
    \includegraphics[width=0.9\textwidth]{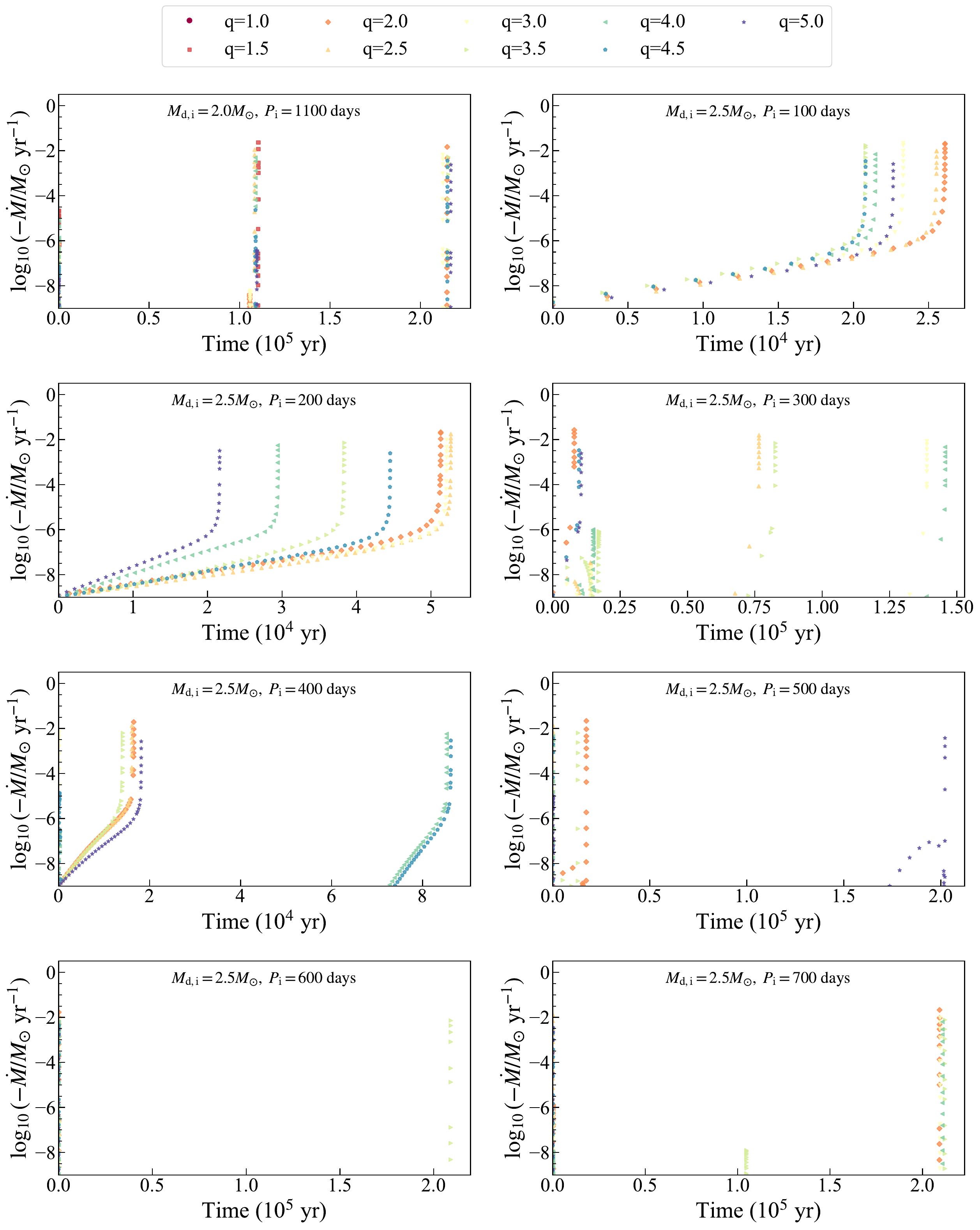}
    \caption{Evolution of the mass transfer rate for all models in our grid. Each panel corresponds to a specific initial donor mass ($M$) and orbital period ($P$). (Continued)}

\end{figure*}
\clearpage

\begin{figure*}[p]
    \ContinuedFloat
    \centering
    \includegraphics[width=0.9\textwidth]{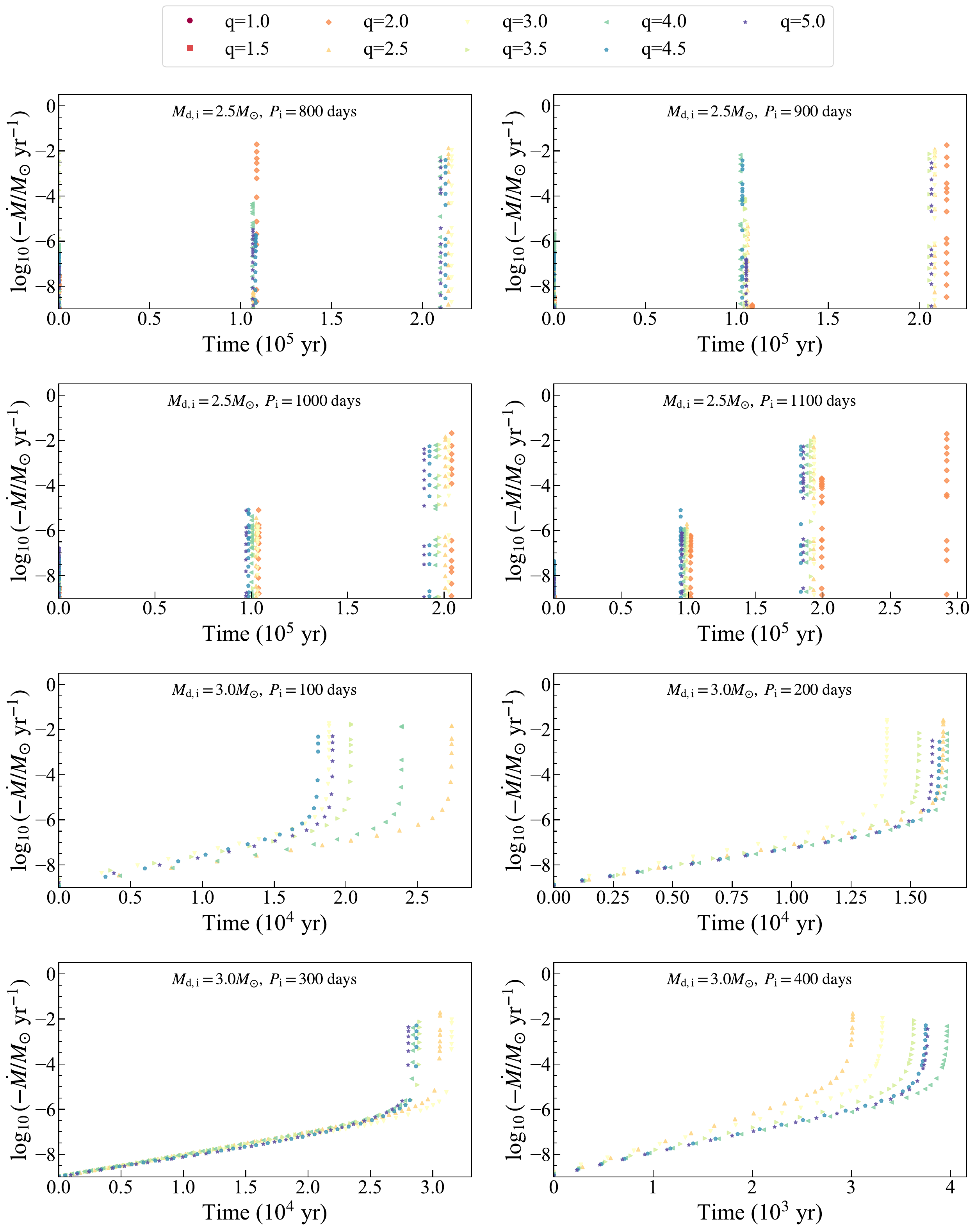}
    \caption{Evolution of the mass transfer rate for all models in our grid. Each panel corresponds to a specific initial donor mass ($M$) and orbital period ($P$). (Continued)}

\end{figure*}
\clearpage

\begin{figure*}[p]
    \ContinuedFloat
    \centering
    \includegraphics[width=0.9\textwidth]{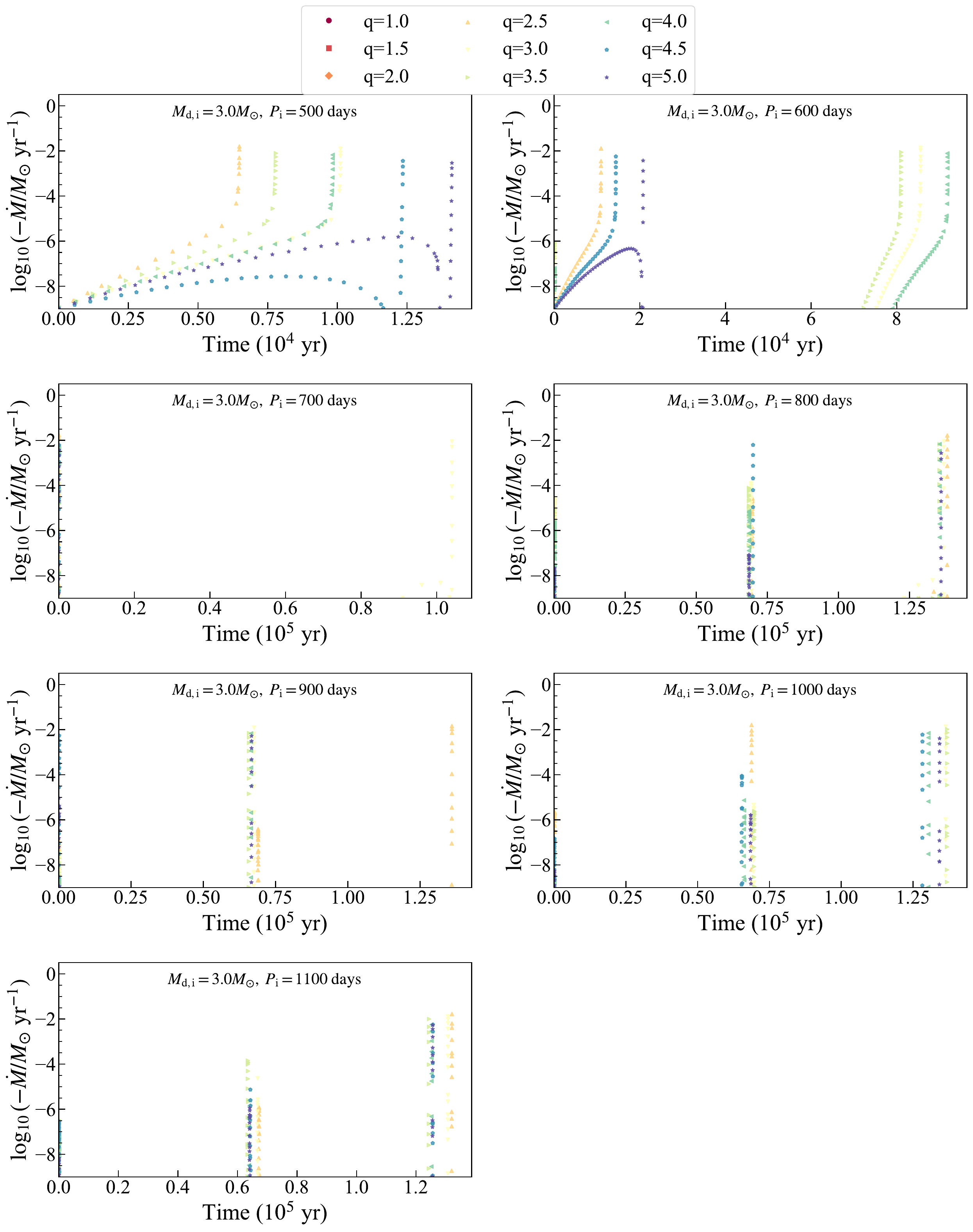}
    \caption{Evolution of the mass transfer rate for all models in our grid. Each panel corresponds to a specific initial donor mass ($M$) and orbital period ($P$). (End)}

\end{figure*}
\clearpage

\end{appendix}

\end{document}